\documentclass[12pt]{article}

\usepackage{amsmath}
\usepackage{graphicx}
\usepackage{cite}
\usepackage{bm}

\newcommand{\ket}[1]{| #1 \rangle}
\newcommand{\bra}[1]{\langle #1 |}
\newcommand{\hcs}[1]{#1^\dagger #1}
\newcommand{\expv}[1]{\langle #1 \rangle}
\begin{document}

\begin{center}
{\Large\bf Local trade-off between information and disturbance
in quantum measurements}
\vskip .6 cm
Hiroaki Terashima
\vskip .4 cm
{\it Department of Science Education, Cooperative Faculty of Education, \\
Gunma University, \\
Maebashi, Gunma 371-8510, Japan}
\vskip .6 cm
\end{center}

\begin{abstract}
This study confirms a local trade-off
between information and disturbance in quantum measurements.
It is represented by the correlation between
the changes in these two quantities
when the measurement is slightly modified.
The correlation indicates that
when the measurement is modified to increase the obtained information,
the disturbance also increases in most cases.
However, the information can be increased while decreasing the disturbance
because the correlation is not necessarily perfect.
For measurements having imperfect correlations,
this paper discusses a general scheme that raises the amount of information
while decreasing the disturbance.
\end{abstract}

\section{Introduction}
An interesting topic in the quantum theory of measurements
is the trade-off between information and disturbance.
In general,
when a measurement provides much information
about the state of a system,
it causes a large disturbance.
This trade-off has been
studied using various measures of
information and
disturbance~\cite{FucPer96,Banasz01,FucJac01,BanDev01,Barnum02,%
DArian03,Ozawa04,GenPar05,MiFiFi05,Maccon06,Sacchi06,BusSac06,Banasz06,%
BuHaHo07,CheLee12,RenFan14,FGNZ15,ShKuUe16}.
Most studies focused on optimal measurements,
which saturate the upper bound of the information for a given disturbance.
An optimal measurement providing more information
always causes a larger disturbance in the system.

By contrast, even for general measurements,
a trade-off can be considered locally.
Consider the neighborhood of a given measurement
in the space of measurements.
Most neighboring measurements
provide more information with larger disturbance
or less information with smaller disturbance.
This means that the information and disturbance
of neighboring measurements are correlated to some extent.
According to this correlation,
when the given measurement is slightly modified
to increase the obtained information,
it usually increases the disturbance in the system.
This is a kind of trade-off between information and disturbance.
It is local in that
it pays attention to
the neighborhood of the given measurement.

However, this trade-off is not strict
because the correlation is imperfect.
The minority of neighboring measurements have the opposite trend.
In particular, unless the given measurement is optimal,
there always exist measurements
that provide more information with smaller disturbances.
This implies that the given measurement can be improved
to enhance the information gain while decreasing the disturbance.
The imperfectness of the correlation
estimates the improvability of the measurement.

This trade-off has not been discussed so far.
Non-optimal measurements are a little interesting
in quantum information theory.
However, in quantum measurement theory,
the trade-off can be utilized to understand
the mathematical structure of measurements.
The correlation clarifies the connections to the neighboring measurements,
giving a geometry in the measurement space.
Moreover, the trade-off can be utilized 
in quantum information processing experiments
to improve realized measurements.
The correlation estimates their improvabilities,
pointing the improvement direction.

This paper mathematically formulates the local trade-off
between information and disturbance
when a measurement is slightly modified.
Suppose that the measurement is performed on
a quantum system in a completely unknown state.
The obtained information and resulting disturbance
are represented as functions of the same parameters.
For these functions,
their directions of steepest ascent and descent
are derived from their gradient vectors.
The angles between these directions quantify
the correlation between the changes in information and disturbance.
They also quantify the improvability
in a general scheme to enhance the measurement.

The remainder of this paper is organized as follows.
Section~\ref{sec:preliminary} covers the review of
information and disturbance
for a single outcome of a quantum measurement.
Section~\ref{sec:steepest} focuses on finding
the steepest-ascent and -descent directions of
the information and disturbance.
Section~\ref{sec:angle} provides the calculations of
the angles between the steepest directions.
Section~\ref{sec:correlation} tackles the correlation
between the information and disturbance.
Section~\ref{sec:improvement} discusses
a general measurement-improvement scheme.
Section~\ref{sec:summary} provides a summary of our results.

\section{\label{sec:preliminary}Preliminaries}
To present a self-contained paper,
we recall the information and disturbance
for a single outcome of a quantum measurement~\cite{Terash15}.
Suppose that a quantum system
is known to be in a pure state $\ket{\psi(a)}$ with probability $p(a)$,
where $a=1,\ldots,N$.
To know the actual state of the system,
we perform a quantum measurement on this system.

A quantum measurement is described by
a set of measurement operators $\{\hat{M}_m\}$ satisfying~\cite{NieChu00}
\begin{equation}
\sum_m\hcs{\hat{M}_m}=\hat{I},
\label{eq:completeness}
\end{equation}
where $\hat{I}$ is identity operator,
and the index $m$ denotes an outcome.
For a system in state $\ket{\psi(a)}$, outcome $m$ is
obtained with probability
\begin{equation}
p(m|a)=\bra{\psi(a)}\hcs{\hat{M}_m}\ket{\psi(a)}.
\end{equation}
After a measurement yielding outcome $m$,
the system's state changes from $\ket{\psi(a)}$ to
\begin{equation}
\ket{\psi(m,a)}=\frac{1}{\sqrt{p(m|a)}}\,\hat{M}_m\ket{\psi(a)}.
\label{eq:postState}
\end{equation}
Note that this measurement
does not change the system's state
to a mixed state.
This type of measurement is called
an ideal measurement~\cite{NieCav97}.

The outcome $m$ provides information on the system's state.
More specifically, from the outcome $m$,
one can naturally estimate the system's state as $\ket{\psi(a'_m)}$,
where $a'_m$ is $a$ that maximizes $p(m|a)$.
The estimation fidelity determines the quality of this estimate, given as
\begin{equation}
 G(m) =\sum_a p(a|m)\,\bigl|\expv{\psi(a'_m)|\psi(a)}\bigr|^2,
\end{equation}
where we have used the conditional
probability of the system's state as $\ket{\psi(a)}$
given outcome $m$:
\begin{equation}
 p(a|m) =\frac{p(m|a)\,p(a)}{p(m)}.
\end{equation}
The total probability of $m$ is
\begin{equation}
 p(m) =\sum_a p(m|a)\,p(a).
\label{eq:totalProb}
\end{equation}
The estimation fidelity $G(m)$ quantifies the information
provided by outcome $m$.

When a measurement yields outcome $m$,
it changes the system's state from $\ket{\psi(a)}$ to $\ket{\psi(m,a)}$
given by Eq.~(\ref{eq:postState}).
This system disturbance can be quantified
by either the size or reversibility of the state change.
The size of the state change can be evaluated
by the operation fidelity
\begin{equation}
 F(m) =\sum_a p(a|m)\bigl|\expv{\psi(a)|\psi(m,a)}\bigr|^2.
\end{equation}
In contrast,
the state change reversibility can be evaluated
on the basis of a reversing measurement~\cite{UeImNa96,Ueda97}.
The reversing measurement reverts the system's state
from $\ket{\psi(m,a)}$ to $\ket{\psi(a)}$ when
it yields a successful outcome.
Using its maximum successful probability~\cite{KoaUed99},
the state change reversibility is evaluated by
the physical reversibility
\begin{equation}
  R(m) = \sum_a p(a|m)\,\frac{\inf_{\ket{\psi}}\,\bra{\psi}\hcs{\hat{M}_m}\ket{\psi}}{p(m|a)}.
\end{equation}
The operation fidelity $F(m)$ and the physical reversibility $R(m)$
quantify the disturbance caused by obtaining outcome $m$.
Both measures decrease as the disturbance increases.

To explicitly calculate $G(m)$, $F(m)$, and $R(m)$,
we assume a completely unknown system state to be measured.
That is, the set of possible states $\{\ket{\psi(a)}\}$ consists
of all pure states of the system,
and $p(a)$ is uniform
according to a normalized invariant measure over the pure states.
In this case, 
the information and disturbance are functions of
the singular values $\{\lambda_{mi}\}$ of $\hat{M}_m$~\cite{Terash15}.
The singular value $\lambda_{mi}$ means the square root of
the probability for obtaining outcome $m$
when the system is in the $i$th eigenstate of
the positive operator-valued measure (POVM)
element $\hat{E}_m=\hat{M}_m^\dagger \hat{M}_m$~\cite{NieChu00}.
Therefore, a measurement
with outcome $m$ can be conveniently expressed
by the $d$-dimensional vector
\begin{equation}
 \bm{\lambda}_m=\left(\lambda_{m1},\lambda_{m2},\ldots,\lambda_{md}\right),
\end{equation}
where $d$ is the Hilbert space dimension of the system.
By definition, the singular values
are not less than $0$.
Moreover, by Eq.~(\ref{eq:completeness}), they cannot exceed $1$.
For simplicity, the singular values are sorted
in the following descending order:
\begin{equation}
 1\ge\lambda_{m1}\ge\lambda_{m2}\ge\cdots\ge\lambda_{md}\ge0,
\label{eq:condLamb}
\end{equation}
where $\lambda_{m1}\neq 0$.

In terms of the singular values,
$G(m)$, $F(m)$, and $R(m)$ are respectively written as~\cite{Terash15}
\begin{align}
 G(m) &= \frac{1}{d+1}\left(1+\frac{\lambda_{m1}^2}{\sigma_m^2}\right),
         \label{eq:Gm} \\
 F(m) &= \frac{1}{d+1}\left(1+\frac{\tau_m^2}{\sigma_m^2}\right),
         \label{eq:Fm} \\
 R(m) &= d\left(\frac{\lambda_{md}^2 }{\sigma_m^2}\right),
         \label{eq:Rm}
\end{align}
where
\begin{equation}
 \sigma_m^2 =\sum_{i=1}^{d}\lambda_{mi}^2, \qquad
 \tau_m =\sum_{i=1}^{d}\lambda_{mi}.
\end{equation}
Note that Eqs.~(\ref{eq:Gm})--(\ref{eq:Rm}) are invariant
under rescaling of the singular values by a constant $c$,
\begin{equation}
 \bm{\lambda}_m \longrightarrow c\bm{\lambda}_m,
\label{eq:rescale}
\end{equation}
and under rearrangement of all singular values
except $\lambda_{m1}$ and $\lambda_{md}$.
In the rearrangement,
$\lambda_{m1}$ is excluded because
it should be the maximum singular value to use Eq.~(\ref{eq:Gm}) for $G(m)$,
and $\lambda_{md}$ is excluded because
it should be the minimum one to use Eq.~(\ref{eq:Rm}) for $R(m)$.

Fundamental measurements are represented
by the following vectors~\cite{Terash16}:
\begin{align}
 \bm{p}^{(d)}_{r} &=c\,(\, \underbrace{1,1,\ldots,1}_{r},
  \underbrace{0,0,\ldots,0}_{d-r}\,), \\
  \bm{m}^{(d)}_{k,l}(\lambda) &=
    c\,(\,\underbrace{1,1,\ldots,1}_{k},
  \underbrace{\lambda,\lambda,\ldots,\lambda}_{l},
  \underbrace{0,0,\ldots,0}_{d-k-l}\,),
\end{align}
where $c$ is a proportionality factor
from the rescaling invariance in Eq.~(\ref{eq:rescale}),
and $\lambda$ is a parameter satisfying $0\le\lambda\le1$.
In particular,
$\bm{p}^{(d)}_{1}$ represents the projective measurement of rank $1$,
achieving the maximum $G(m)$ with the minimum $F(m)$ and $R(m)$.
Conversely,
$\bm{p}^{(d)}_{d}$ represents the identity operation,
achieving the minimum $G(m)$ with the maximum $F(m)$ and $R(m)$.
Moreover, the measurements represented by
$\bm{m}^{(d)}_{1,d-1}(\lambda)$ are the optimal measurements,
saturating the upper bounds of $G(m)$
for given $F(m)$ or $R(m)$~\cite{Terash16}.

\section{\label{sec:steepest}Steepest Directions}
Herein, we find the directions of steepest ascent and descent
of $G(m)$, $F(m)$, and $R(m)$.
Under the conditions of Eq.~(\ref{eq:condLamb}),
these directions are not necessarily parallel or antiparallel to
the gradient vectors of $G(m)$, $F(m)$, and $R(m)$.
That is, consider modifying a measurement $\bm{\lambda}_m$
by an infinitesimal vector $\bm{\epsilon}_m$ as
\begin{equation}
 \bm{\lambda}'_m = \bm{\lambda}_m +\bm{\epsilon}_m.
\label{eq:modification}
\end{equation}
However, $\bm{\epsilon}_m$ cannot be arbitrary.
After the modification,
$\bm{\lambda}'_m$ must also satisfy
the conditions of Eq.~(\ref{eq:condLamb}).

\begin{figure}
\begin{center}
\includegraphics[scale=0.42]{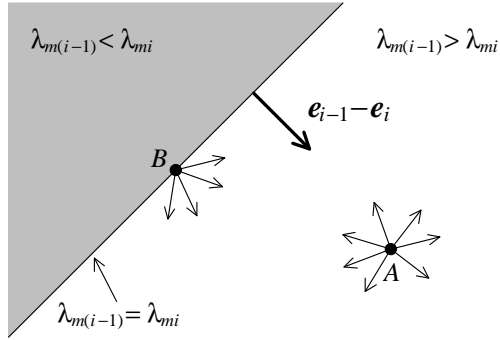}
\end{center}
\caption{\label{fig1}
Measurement and its modification.
The points $A$ and $B$ denote measurements,
and the attached arrows connote
some possible modifications.
The gray region shows a forbidden region $\lambda_{m(i-1)}<\lambda_{mi}$.
The vector $\bm{e}_{i-1}-\bm{e}_i$ is normal to
its boundary $\lambda_{m(i-1)}=\lambda_{mi}$.}
\end{figure}%
Figure~\ref{fig1} shows a sketch of this situation.
Although the measurement $A$ can accept any modification,
the measurement $B$ cannot because
some modifications would move it into the gray region
forbidden by Eq.~(\ref{eq:condLamb}).
As long as the modification is infinitesimal,
such a violation occurs only when
the measurement is on any of the boundaries.
Therefore, $\bm{\epsilon}_m$ is restricted
when $\bm{\lambda}_m$ has some equal signs in Eq.~(\ref{eq:condLamb}).

However, not all inequalities in Eq.~(\ref{eq:condLamb}) are relevant.
The relevant inequalities are
$\lambda_{md}\ge 0$, $\lambda_{m1}\ge \lambda_{mi}$,
and $\lambda_{mi}\ge \lambda_{md}$.
The other inequalities can be ignored
by rescaling and rearranging $\bm{\lambda}'_m$.
For example, if $\lambda'_{m1}>1$,
$\bm{\lambda}'_m$ is rescaled by Eq.~(\ref{eq:rescale})
to satisfy $\lambda'_{m1}\le 1$,
and if $\lambda'_{m2}<\lambda'_{m3}$,
$\lambda'_{m2}$ and $\lambda'_{m3}$ are interchanged
to satisfy $\lambda'_{m2}>\lambda'_{m3}$.
Note that these operations do not affect $G(m)$, $F(m)$, and $R(m)$.
To keep the three relevant inequalities,
$\bm{\epsilon}_m$ should satisfy
\begin{align}
  \bm{e}_d \cdot \bm{\epsilon}_m &\ge 0
       \qquad \text{if $\lambda_{md}=0$},
       \label{eq:condEps1} \\
  \left(\bm{e}_1-\bm{e}_i\right) \cdot \bm{\epsilon}_m &\ge 0
      \qquad \text{if $\lambda_{m1}=\lambda_{mi}$},
       \label{eq:condEps2}  \\
  \left(\bm{e}_i-\bm{e}_d\right) \cdot \bm{\epsilon}_m &\ge 0
      \qquad \text{if $\lambda_{mi}=\lambda_{md}$},
       \label{eq:condEps3} 
\end{align}
where $\bm{e}_i$ is the unit vector along the $i$th axis.
These are because
$\bm{e}_d$, $\bm{e}_1-\bm{e}_i$, and $\bm{e}_i-\bm{e}_d$ are normal
to the boundaries $\lambda_{md}=0$, $\lambda_{m1}=\lambda_{mi}$,
and $\lambda_{mi}=\lambda_{md}$,
respectively (see Fig.~\ref{fig1}).

Under the conditional
equations (\ref{eq:condEps1})--(\ref{eq:condEps3}),
we consider the steepest directions of $G(m)$, $F(m)$, and $R(m)$.
As shown in Appendix~\ref{sec:vector},
they are derived from the gradient vectors
$\bm{\nabla} G(m)$, $\bm{\nabla} F(m)$, and $\bm{\nabla} R(m)$
in Eqs.~(\ref{eq:gradG})--(\ref{eq:gradR}).
Three unit vectors are obtained for each function as follows.

The first vector is a unit vector
in the gradient direction.
For $G(m)$, $F(m)$, and $R(m)$,
it is given by
\begin{align}
  \bm{g}_m
   &=\frac{\sigma_m}{\sqrt{\sigma_m^2-\lambda_{m1}^2}}
       \left(\bm{e}_1
      -\frac{\lambda_{m1}}{\sigma_m^2}\bm{\lambda}_m\right),
            \label{eq:unitG} \\
  \bm{f}_m
   &=\frac{\sigma_m}{\sqrt{d\sigma_m^2-\tau_m^2}}
       \left(\bm{l}_d-\frac{\tau_m}{\sigma_m^2}\bm{\lambda}_m\right),
            \label{eq:unitF} \\
  \bm{r}_m
   &=\frac{\sigma_m}{\sqrt{\sigma_m^2-\lambda_{md}^2}}
       \left(\bm{e}_d
      -\frac{\lambda_{md}}{\sigma_m^2}\bm{\lambda}_m\right),
           \label{eq:unitR}
\end{align}
respectively, where
\begin{equation}
 \bm{l}_{n}=\sum_{i=1}^{n}\bm{e}_i.
\label{eq:vecLn}
\end{equation}

\begin{figure}
\begin{center}
\includegraphics[scale=0.42]{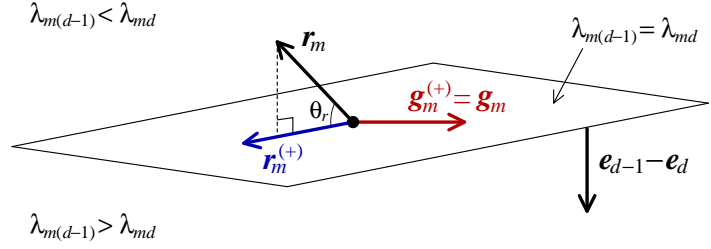}
\end{center}
\caption{\label{fig2}
Directions of gradient and steepest ascent on boundary.
The plane denotes the boundary $\lambda_{m(d-1)}=\lambda_{md}$
with the normal vector $\bm{e}_{d-1}-\bm{e}_d$,
and the region above it
is the forbidden region $\lambda_{m(d-1)}<\lambda_{md}$.
The vectors $\bm{r}_m$ and $\bm{r}^{(+)}_m$
are the unit vectors of $R(m)$
in the gradient and steepest-ascent directions,
respectively, for a measurement on the boundary.
The vectors $\bm{g}_m$ and $\bm{g}^{(+)}_m$
are unit vectors of $G(m)$.
The angle between $\bm{r}^{(+)}_m$ and $\bm{r}_m$ is $\theta_r$.}
\end{figure}%
The second vector is a unit vector
in the steepest-ascent direction.
For $G(m)$, $F(m)$, and $R(m)$,
it is given by
\begin{align}
 \bm{g}^{(+)}_m &= \bm{g}_m, \label{eq:ascG} \\
 \bm{f}^{(+)}_m &= \bm{f}_m, \label{eq:ascF} \\
 \bm{r}^{(+)}_m
   &=  \frac{\sqrt{n_d}\,\sigma_m}{\sqrt{\sigma_m^2-n_d\lambda_{md}^2}}
            \left[\frac{1}{n_d}\left(\bm{l}_{d}-\bm{l}_{d-n_d}\right)
            -\frac{\lambda_{md}}{\sigma_m^2}\bm{\lambda}_m\right],
           \label{eq:ascR}
\end{align}
respectively,
where $n_d$ is the degeneracy of the minimum singular value.
If the minimum singular value degenerates as $n_d\neq1$
(e.g., $\lambda_{m(d-1)}=\lambda_{md}$ with $n_d=2$),
$\bm{r}^{(+)}_m$ differs from $\bm{r}_m$.
In this case,
$\bm{r}_m$ points
from the boundary $\lambda_{m(d-1)}=\lambda_{md}$
into the forbidden region $\lambda_{m(d-1)}<\lambda_{md}$
as illustrated in Fig.~\ref{fig2},
violating the condition of Eq.~(\ref{eq:condEps3}).
The steepest-ascent direction $\bm{r}^{(+)}_m$ is
obtained by projecting $\bm{r}_m$ onto the boundary
and normalizing the projected vector to length $1$
(see Appendix~\ref{sec:vector}).
The difference is 
the angle $\theta_r$
measured between $\bm{r}^{(+)}_m$ and $\bm{r}_m$,
\begin{equation}
 \cos\theta_r =\bm{r}^{(+)}_m\cdot \bm{r}_m=
      \sqrt{\frac{\sigma_m^2-n_d\lambda_{md}^2}
     {n_d\left(\sigma_m^2-\lambda_{md}^2\right)}}.
\label{eq:cosR}
\end{equation}

The third vector is a unit vector
in the steepest-descent direction.
For $G(m)$, $F(m)$, and $R(m)$,
it is given by
\begin{align}
  \bm{g}^{(-)}_m
    &= -\frac{\sqrt{n_1}\,\sigma_m}{\sqrt{\sigma_m^2-n_1\lambda_{m1}^2}}
       \left(\frac{1}{n_1}\bm{l}_{n_1}
      -\frac{\lambda_{m1}}{\sigma_m^2}\bm{\lambda}_m\right),
            \label{eq:dscG} \\
  \bm{f}^{(-)}_m
   &=-\frac{\sigma_m}{\sqrt{\left(d-n_0\right)\sigma_m^2-\tau_m^2}}
        \left(\bm{l}_{d-n_0}
         -\frac{\tau_m}{\sigma_m^2}\bm{\lambda}_m\right),
            \label{eq:dscF} \\
  \bm{r}^{(-)}_m
   &= -\delta_{n_0,0}\,\bm{r}_m,
           \label{eq:dscR}
\end{align}
respectively,
where $n_1$ is the degeneracy of the maximum singular value, and
$n_0$ is that of the singular value $0$.
If the maximum singular value degenerates as $n_1\neq1$
(e.g., $\lambda_{m1}=\lambda_{m2}$ with $n_1=2$),
$\bm{g}^{(-)}_m$ differs from $-\bm{g}_m$.
The difference is 
the angle $\theta_g$
measured between $\bm{g}^{(-)}_m$ and $-\bm{g}_m$,
\begin{equation}
 \cos\theta_g =-\bm{g}^{(-)}_m\cdot \bm{g}_m
   =\sqrt{\frac{\sigma_m^2-n_1\lambda_{m1}^2}
               {n_1\left(\sigma_m^2-\lambda_{m1}^2\right)}}.
\label{eq:cosG}
\end{equation}
Similarly,
if some singular values are $0$ as $n_0\neq0$
(e.g., $\lambda_{md}=0$ with $n_0=1$),
$\bm{f}^{(-)}_m$ differs from $-\bm{f}_m$.
The difference is 
the angle $\theta_f$
measured between $\bm{f}^{(-)}_m$ and $-\bm{f}_m$,
\begin{equation}
 \cos\theta_f=-\bm{f}^{(-)}_m\cdot \bm{f}_m
   =\sqrt{\frac{\left(d-n_0\right)\sigma_m^2-\tau_m^2}
                 {d\sigma_m^2-\tau_m^2}}.
\label{eq:cosF}
\end{equation}
In this case,
$\bm{r}^{(-)}_m$
is not $-\bm{r}_m$ but a zero vector $\bm{0}$.

The above equations give $\bm{0}/0$ or $0/0$,
when a zero vector is normalized.
For example, when $\bm{\lambda}_m=\bm{p}^{(d)}_{1}$,
Eq.~(\ref{eq:unitG}) gives $\bm{g}_m=\bm{0}/0$
because $\bm{\nabla} G(m)=\bm{0}$ from Eq.~(\ref{eq:gradG})
in Appendix~\ref{sec:vector}.
Unfortunately, the limit of $\bm{g}_m$ as
$\bm{\lambda}_m\to\bm{p}^{(d)}_{1}$ does not exist.
In such cases, we simply assume $\bm{0}/0=\bm{0}$ and $0/0=0$.
Specifically,
\begin{align}
\bm{g}_m=\bm{g}^{(+)}_m = \bm{0}, \quad 
\cos\theta_g =0 &\quad
 \mbox{at $\bm{\lambda}_m=\bm{p}^{(d)}_{1}$}, \label{eq:zeroVec1} \\
\bm{f}_m=\bm{f}^{(+)}_m=\bm{r}^{(+)}_m = \bm{0}, \quad 
\cos\theta_f=0 &\quad 
 \mbox{at $\bm{\lambda}_m=\bm{p}^{(d)}_{d}$}, \label{eq:zeroVec2} \\
\bm{g}^{(-)}_m=\bm{f}^{(-)}_m = \bm{0} &\quad 
 \mbox{at $\bm{\lambda}_m=\bm{p}^{(d)}_{r}$}. \label{eq:zeroVec3}
\end{align}

\section{\label{sec:angle}Angles}
In this section, we consider the angle between the steepest directions
of the information and disturbance.
This angle concerns the local trade-off between 
the information and disturbance 
when a measurement is slightly modified.
Two information--disturbance pairs are discussed herein:
$G(m)$ versus $F(m)$ and $G(m)$ versus $R(m)$.

\begin{figure}
\begin{center}
\includegraphics[scale=0.42]{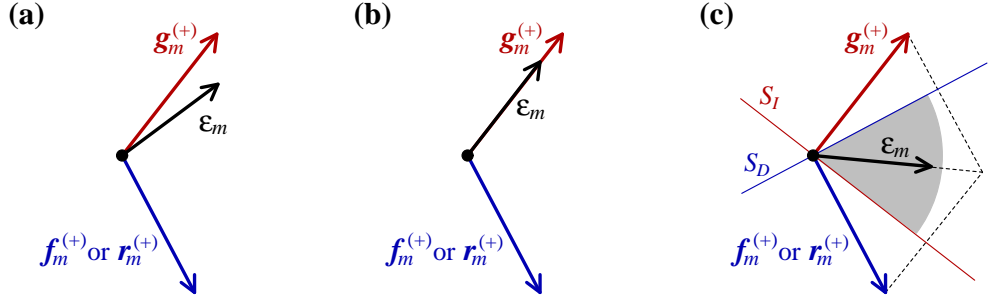}
\end{center}
\caption{\label{fig3}
Relation between angle and trade-off.
The vectors $\bm{g}^{(+)}_m$, $\bm{f}^{(+)}_m$, and $\bm{r}^{(+)}_m$
are unit vectors
in the steepest-ascent directions of $G(m)$, $F(m)$, and $R(m)$,
respectively.
The vector $\bm{\epsilon}_m$ is a measurement modification.
In panel (c),
the line $S_I$ denotes constant-information surface according to $G(m)$,
whereas the line $S_D$ denotes
constant-disturbance surface according to $F(m)$ or $R(m)$.}
\end{figure}%
Figure~\ref{fig3} illustrates the relation between
the angle and the trade-off.
Suppose that the measurement is modified by
$\bm{\epsilon}_m$ as in Eq.~(\ref{eq:modification}).
To effectively increase the obtained information,
$\bm{\epsilon}_m$ should align in a close direction to $\bm{g}^{(+)}_m$,
as shown in Fig.~\ref{fig3}(a).
However, in this case, $\bm{\epsilon}_m$ points
in an approximately opposite direction to
$\bm{f}^{(+)}_m$ or $\bm{r}^{(+)}_m$,
because the angles between $\bm{g}^{(+)}_m$ and $\bm{f}^{(+)}_m$,
and $\bm{g}^{(+)}_m$ and $\bm{r}^{(+)}_m$
are usually obtuse as will be shown later.
This means that
such a modification usually increases the disturbance in the system
as a trade-off,
since decreasing $F(m)$ or $R(m)$ means increasing disturbance.
The wider the angle, the larger the trade-off.

For $G(m)$ versus $F(m)$,
the angle between $\bm{g}^{(+)}_m$ and $\bm{f}^{(+)}_m$
is expressed by their dot product
$C^{(++)}_{GF}=\bm{g}^{(+)}_m\cdot \bm{f}^{(+)}_m$.
This dot product is the cosine of the angle
because the two vectors are normalized.
Its value is determined as
\begin{equation}
  C^{(++)}_{GF}
      =-\frac{\tau_m\lambda_{m1}-\sigma_m^2}
      {\sqrt{\left(\sigma_m^2-\lambda_{m1}^2\right)
       \left(d\sigma_m^2-\tau_m^2\right)}}\le 0
\label{eq:CGF}
\end{equation}
from Eqs.~(\ref{eq:unitG}), (\ref{eq:unitF}),
(\ref{eq:ascG}), and (\ref{eq:ascF}).
Note that by assuming $0/0=0$, $C^{(++)}_{GF}=0$
if $\bm{\lambda}_m=\bm{p}^{(d)}_{1}$
or $\bm{\lambda}_m=\bm{p}^{(d)}_{d}$.
The last inequality of Eq.~(\ref{eq:CGF}) can be proven as
\begin{equation}
  \tau_m\lambda_{m1}-\sigma_m^2=
   \sum_i \lambda_{mi} (\lambda_{m1}-\lambda_{mi}) \ge 0.
\end{equation}
As $C^{(++)}_{GF}\le0$, the angle
between $\bm{g}^{(+)}_m$ and $\bm{f}^{(+)}_m$ is either right or obtuse.
The maximum value $C^{(++)}_{GF}=0$ is achieved at
$\bm{\lambda}_m=\bm{p}^{(d)}_{r}$, whereas
the minimum value $C^{(++)}_{GF}=-1$ is achieved at
optimal measurements $\bm{\lambda}_m=\bm{m}^{(d)}_{1,d-1}(\lambda)$.

\begin{figure}
\begin{center}
\includegraphics[scale=0.52]{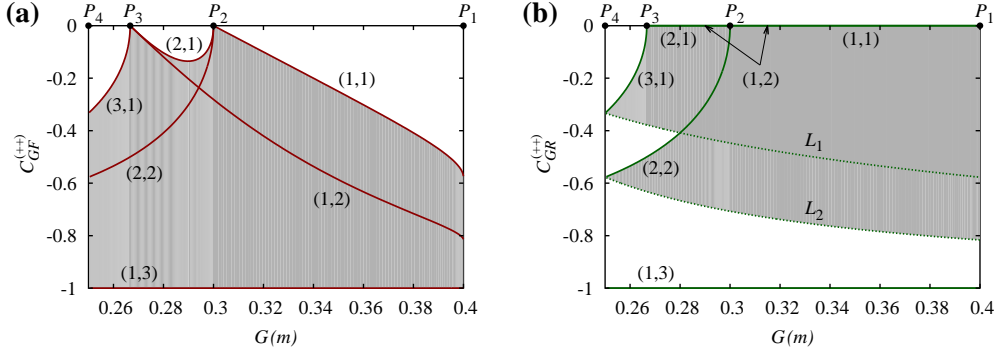}
\end{center}
\caption{\label{fig4}
Ranges of angles between
(a) $\bm{g}^{(+)}_m$ and $\bm{f}^{(+)}_m$, and
(b) $\bm{g}^{(+)}_m$ and $\bm{r}^{(+)}_m$.
The gray regions show the possible ranges of
$C^{(++)}_{GF}=\bm{g}^{(+)}_m\cdot \bm{f}^{(+)}_m$
and $C^{(++)}_{GR}=\bm{g}^{(+)}_m\cdot \bm{r}^{(+)}_m$
as functions of $G(m)$ in $d=4$.
The point $P_r$ denotes $\bm{p}^{(d)}_{r}$,
the line $(k,l)$ denotes $\bm{m}^{(d)}_{k,l}(\lambda)$,  and
the dotted line $L_n$ in panel (b) denotes
the limits as measurements having $n_d=n$ approach the optimal ones.}
\end{figure}%
Figure~\ref{fig4}(a) shows the possible range of $C^{(++)}_{GF}$
as a function of $G(m)$ in $d=4$.
The possible range is determined similarly to
Appendix~A of Ref.~\cite{Terash16}.
The point $P_r$ denotes $\bm{p}^{(d)}_{r}$,
and the line $(k,l)$ denotes $\bm{m}^{(d)}_{k,l}(\lambda)$
as $0<\lambda<1$.
$C^{(++)}_{GF}$ cannot have unique limits
as $\bm{\lambda}_m\to\bm{p}^{(d)}_{1}$
and $\bm{\lambda}_m\to\bm{p}^{(d)}_{d}$,
where $C^{(++)}_{GF}=0$
by Eqs.~(\ref{eq:zeroVec1}) and (\ref{eq:zeroVec2}).
For example,
although both $\bm{m}^{(4)}_{3,1}(\lambda)$
and $\bm{m}^{(4)}_{2,2}(\lambda)$
become $\bm{p}^{(4)}_{4}$ at $\lambda=1$,
they give different limits of $C^{(++)}_{GF}$ as $\lambda\to1$.
Figure~\ref{fig4}(a) shows that
their corresponding lines $(3,1)$ and $(2,2)$ do not coincide at
the left ends as $\lambda\to1$, i.e., at $G(m)=0.25$.

Similarly, for $G(m)$ versus $R(m)$,
the cosine of the angle between $\bm{g}^{(+)}_m$ and $\bm{r}^{(+)}_m$
is given by $C^{(++)}_{GR}=\bm{g}^{(+)}_m\cdot \bm{r}^{(+)}_m$,
determined as
\begin{equation}
  C^{(++)}_{GR}
     =-\frac{\sqrt{n_d}\,\lambda_{m1}\lambda_{md}
                       \left(1-\delta_{n_d,d}\right)}
             {\sqrt{\left(\sigma_m^2-\lambda_{m1}^2\right)
                    \left(\sigma_m^2-n_d\lambda_{md}^2\right)}}
      \le 0
\label{eq:CGR}
\end{equation}
from Eqs.~(\ref{eq:unitG}), (\ref{eq:ascG}), and (\ref{eq:ascR}).
Note that by assuming $0/0=0$, $C^{(++)}_{GR}=0$
if $\bm{\lambda}_m=\bm{p}^{(d)}_{1}$
or $\bm{\lambda}_m=\bm{p}^{(d)}_{d}$,
where $n_d=d$ when $\bm{\lambda}_m=\bm{p}^{(d)}_{d}$.
The angle between $\bm{g}^{(+)}_m$ and $\bm{r}^{(+)}_m$
is also either right or obtuse.
The maximum value $C^{(++)}_{GR}=0$ is achieved at
$\lambda_{md}=0$ or $\bm{\lambda}_m=\bm{p}^{(d)}_{d}$,
whereas the minimum value $C^{(++)}_{GR}=-1$ is achieved at
optimal measurements
$\bm{\lambda}_m=\bm{m}^{(d)}_{1,d-1}(\lambda)$ using $n_d=d-1$.

Figure~\ref{fig4}(b) shows the possible range of $C^{(++)}_{GR}$
as a function of $G(m)$ in $d=4$.
The dotted line $L_n$ denotes
the limits of $C^{(++)}_{GR}$
as measurements having $n_d=n$ approach the optimal ones,
that is, the values obtained for
$\bm{\lambda}_m=\bm{m}^{(d)}_{1,d-1}(\lambda)$ but using $n_d=n$
instead of $n_d=d-1$.
This line is an open boundary because $C^{(++)}_{GR}$ jumps to $-1$
at $\bm{\lambda}_m=\bm{m}^{(d)}_{1,d-1}(\lambda)$.
By the lines $(k,l)$ and $L_n$,
the region is divided into overlapping subregions according to $n_d$.
Similar to the case of $C^{(++)}_{GF}$,
$C^{(++)}_{GR}$ cannot have unique limits
as $\bm{\lambda}_m\to\bm{p}^{(d)}_{1}$
and $\bm{\lambda}_m\to\bm{p}^{(d)}_{d}$,
where $C^{(++)}_{GR}=0$
by Eqs.~(\ref{eq:zeroVec1}) and (\ref{eq:zeroVec2}).

The above angles are compared with the angles between the gradient vectors,
$C_{GF}=\bm{g}_m\cdot \bm{f}_m$ and $C_{GR}=\bm{g}_m\cdot \bm{r}_m$.
From Eqs.~(\ref{eq:ascG}) and (\ref{eq:ascF}),
\begin{equation}
C_{GF}=C^{(++)}_{GF}.
\label{eq:CGF0}
\end{equation}
However, $\bm{r}_m$ is not equal to $\bm{r}^{(+)}_m$
if $n_d>1$, as in Eqs.~(\ref{eq:unitR}) and (\ref{eq:ascR}).
Hence, $C_{GR}$ is different from $C^{(++)}_{GR}$,
given by
\begin{equation}
  C_{GR} =-\frac{\lambda_{m1}\lambda_{md}}
             {\sqrt{\left(\sigma_m^2-\lambda_{m1}^2\right)
                    \left(\sigma_m^2-\lambda_{md}^2\right)}} \le 0.
\label{eq:CGR0}
\end{equation}
From Eq.~(\ref{eq:cosR}), 
$C^{(++)}_{GR}$ and $C_{GR}$ are related as
\begin{equation}
 C_{GR}=C^{(++)}_{GR}\cos\theta_r
\label{eq:CGR2CGR}
\end{equation}
if $\bm{\lambda}_m\neq\bm{p}^{(d)}_{d}$.
However, if $\bm{\lambda}_m=\bm{p}^{(d)}_{d}$,
$C^{(++)}_{GR}=\cos\theta_r=0$ but $C_{GR}=-1/(d-1)$.

Any measurement cannot achieve $C_{GR}=-1$.
Even the optimal measurements $\bm{m}^{(d)}_{1,d-1}(\lambda)$
give $C_{GR}>-1$ with $\lambda$-dependence.
As Fig.~\ref{fig2} illustrates,
if $n_d>1$ like the optimal measurements,
$\bm{g}_m$ lies on the boundary
$\lambda_{m(d-1)}=\lambda_{md}$
because $\left(\bm{e}_{d-1}-\bm{e}_d\right) \cdot \bm{g}_m=0$,
whereas $\bm{r}_m$ does not.
Therefore,
$\bm{r}_m$ cannot be antiparallel to $\bm{g}_m$.
In contrast, $\bm{r}^{(+)}_m$ can because
it is obtained by projecting $\bm{r}_m$ onto the boundary.

Using $C^{(++)}_{GF}$ and $C^{(++)}_{GR}$,
we now discuss an example of a local trade-off
between information and disturbance.
When a measurement $\bm{\lambda}_m$ is modified
by $\bm{\epsilon}_m$ as in Eq.~(\ref{eq:modification}),
$G(m)$, $F(m)$, and $R(m)$ respectively change as follows:
\begin{align}
 \Delta G(m) &= \bm{\epsilon}_m\cdot \bm{\nabla} G(m)
         =\left(\bm{\epsilon}_m\cdot \bm{g}_m\right)
                \left\| \bm{\nabla} G(m) \right\| , 
 \label{eq:changeG} \\
 \Delta F(m) &= \bm{\epsilon}_m\cdot \bm{\nabla} F(m)
         =\left(\bm{\epsilon}_m\cdot \bm{f}_m\right)
                \left\| \bm{\nabla} F(m) \right\|,
 \label{eq:changeF} \\
 \Delta R(m) &= \bm{\epsilon}_m\cdot \bm{\nabla} R(m)
            =\left(\bm{\epsilon}_m\cdot \bm{r}_m\right)
                \left\| \bm{\nabla} R(m) \right\|.
\label{eq:changeR}
\end{align}
As an example, $\bm{\epsilon}_m$ is set to
$\epsilon\bm{g}^{(+)}_m$ with a positive infinitesimal $\epsilon$ to
increase $G(m)$ as far as possible, as shown in Fig.~\ref{fig3}(b).
In this case, $G(m)$ increases by
\begin{equation}
\left[\Delta G(m)\right]_\mathrm{max}
   =\epsilon\left\| \bm{\nabla} G(m) \right\| \ge 0.
\label{eq:maxDelG}
\end{equation}
However, $F(m)$ decreases as
\begin{equation}
 \Delta F(m) =C^{(++)}_{GF}\left[\Delta F(m)\right]_\mathrm{max}\le 0,
\label{eq:tradeGF}
\end{equation}
where
\begin{equation}
\left[\Delta F(m)\right]_\mathrm{max}
   =\epsilon\left\| \bm{\nabla} F(m) \right\| \ge 0
\label{eq:maxDelF}
\end{equation}
is $\Delta F(m)$ when $F(m)$ is increased as far as possible
by $\bm{\epsilon}_m=\epsilon\bm{f}^{(+)}_m$.
Similarly, $R(m)$ decreases as
\begin{equation}
 \Delta R(m)=C^{(++)}_{GR}\left[\Delta R(m)\right]_\mathrm{max}\le0
\label{eq:tradeGR}
\end{equation}
from Eq.~(\ref{eq:CGR2CGR}),
where
\begin{equation}
 \left[\Delta R(m)\right]_\mathrm{max}
  =\epsilon\left\| \bm{\nabla} R(m) \right\|\cos\theta_r\ge0
\label{eq:maxDelR}
\end{equation}
is $\Delta R(m)$ when $R(m)$ is increased as far as possible
by $\bm{\epsilon}_m=\epsilon\bm{r}^{(+)}_m$
(not by $\bm{\epsilon}_m=\epsilon\bm{r}_m$).
Note that
$\bm{\epsilon}_m=\epsilon\bm{r}_m$ is forbidden
under the condition of Eq.~(\ref{eq:condEps3}).
Equation~(\ref{eq:tradeGF}) shows a local trade-off
between $G(m)$ and $F(m)$, and
Eq.~(\ref{eq:tradeGR}) shows that
between $G(m)$ and $R(m)$.

These trade-offs,
described by $C^{(++)}_{GF}$ and $C^{(++)}_{GR}$,
are special cases obtained for
$\bm{\epsilon}_m=\epsilon\bm{g}^{(+)}_m$.
To describe the entire local trade-off,
the angles between the other steepest directions,
such as $C^{(--)}_{GF}=\bm{g}^{(-)}_m\cdot \bm{f}^{(-)}_m$,
are also needed, as will be shown in the next section.
The remaining angles are summarized
in Appendix~\ref{sec:cosine}.

\section{\label{sec:correlation}Correlation}
To describe the entire local trade-off,
we consider the correlation between the information and disturbance changes.
The two changes are plotted on an information--disturbance plane
for various measurement modifications.
The points plotted are distributed in a region
characterized by four different angles between
the steepest directions of the information and disturbance.

For $G(m)$ versus $F(m)$,
when a measurement $\bm{\lambda}_m$ is modified by $\bm{\epsilon}_m$,
the changes of the information and the disturbance
are given by Eqs.~(\ref{eq:changeG}) and (\ref{eq:changeF}), respectively.
They are normalized as
\begin{align}
  \Delta g_m &=\frac{\Delta G(m)}{\left\| \bm{\epsilon}_m \right\|
                                  \left\|\bm{\nabla} G(m) \right\|}
              =\frac{\bm{\epsilon}_m}{\left\| \bm{\epsilon}_m \right\|}
               \cdot \bm{g}_m, 
         \label{eq:normalChangeG} \\
  \Delta f_m &=\frac{\Delta F(m)}{\left\| \bm{\epsilon}_m \right\|
                                  \left\|\bm{\nabla} F(m) \right\|}
              =\frac{\bm{\epsilon}_m}{\left\| \bm{\epsilon}_m \right\|}
               \cdot \bm{f}_m,
         \label{eq:normalChangeF}
\end{align}
to make it easier to compare different measurements.

%\begin{figure*}
\begin{figure}
\begin{center}
\includegraphics[scale=0.49]{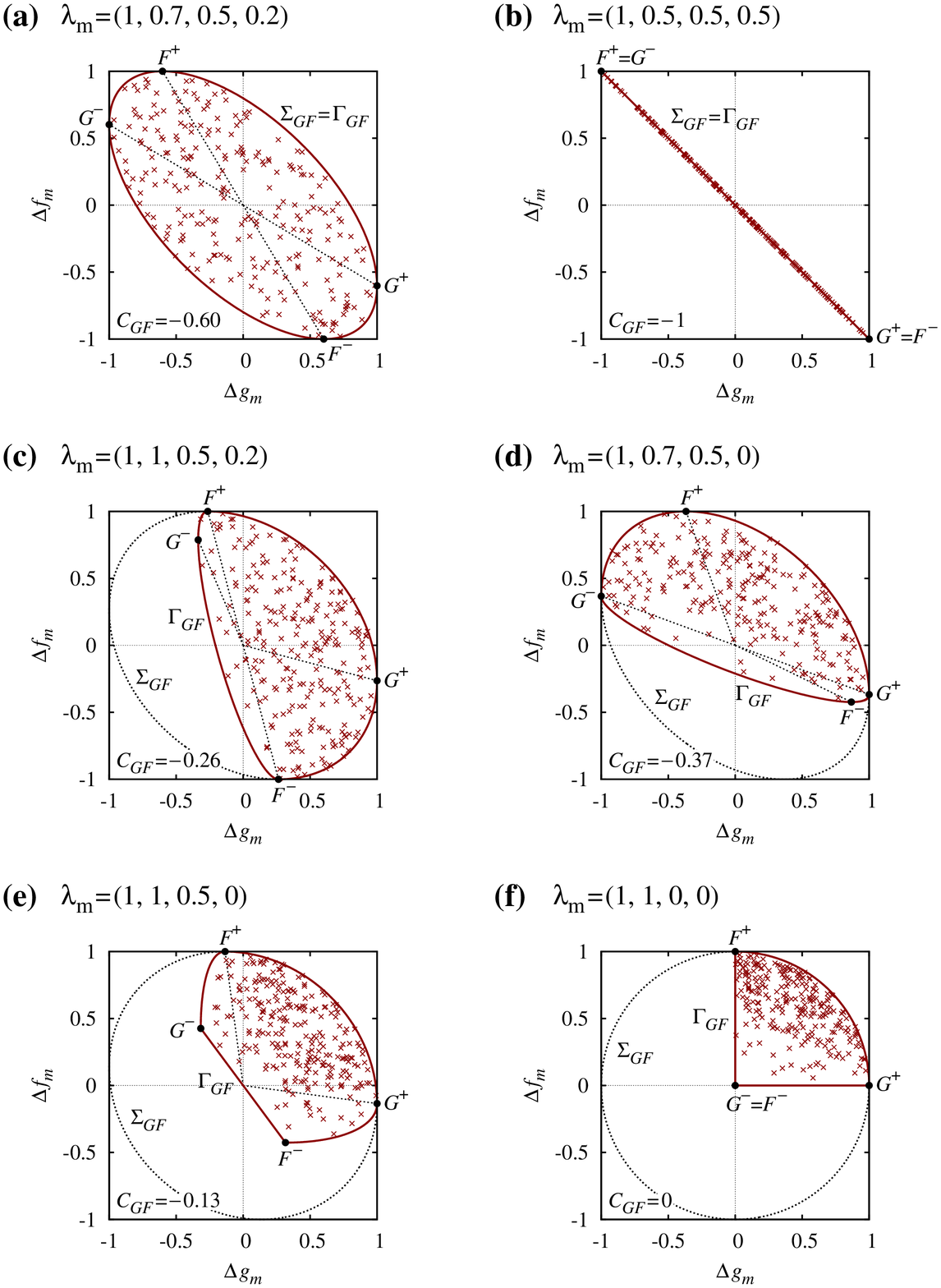}
\end{center}
\caption{\label{fig5}
Correlation between changes in $G(m)$ and $F(m)$.
For six different measurements of $d=4$,
the normalized changes $\Delta g_m$ and $\Delta f_m$
are plotted using $250$ random $\bm{\epsilon}_m$.
The ellipse $\Sigma_{GF}$, characterized by $C_{GF}$, encloses
the region if $\bm{\epsilon}_m$'s were unconditional,
whereas the boundary $\Gamma_{GF}$ encloses
the region when $\bm{\epsilon}_m$'s are
conditioned by Eqs.~(\ref{eq:condEps1}) and (\ref{eq:condEps2}).
The points $G^\pm$ and $F^\pm$ are given by
$\bm{\epsilon}_m=\epsilon\bm{g}^{(\pm)}_m$ and
$\bm{\epsilon}_m=\epsilon\bm{f}^{(\pm)}_m$, respectively.
}
\end{figure}%
%\end{figure*}%
For various $\bm{\epsilon}_m$,
$\Delta g_m$ and $\Delta f_m$ are plotted on a plane.
The modification $\bm{\epsilon}_m$ should satisfy
the conditions of Eqs.~(\ref{eq:condEps1}) and (\ref{eq:condEps2}).
However, there is no need to impose 
the condition of Eq.~(\ref{eq:condEps3}) on $\bm{\epsilon}_m$
for the case of $G(m)$ versus $F(m)$.
This is because as long as $R(m)$ is not used,
the inequality $\lambda_{mi}\ge \lambda_{md}$ in Eq.~(\ref{eq:condLamb})
can also be ignored by rearranging $\bm{\lambda}'_m$.
Figure~\ref{fig5} shows the plotted graphs
for six measurements of $d=4$.
The points were generated
by $250$ random $\bm{\epsilon}_m$
normalized as $\left\| \bm{\epsilon}_m \right\|=0.01$.

If $n_1=1$ and $n_0=0$,
the measurement $\bm{\lambda}_m$ is away from
the relevant boundaries $\lambda_{m1}=\lambda_{mi}$ and $\lambda_{md}=0$,
like the measurement $A$ in Fig.~\ref{fig1}.
Therefore,
it can accept any $\bm{\epsilon}_m$
without constraint from Eqs.~(\ref{eq:condEps1}) and (\ref{eq:condEps2}).
Figure~\ref{fig5}(a) shows this case.
The plotted points lie inside the ellipse $\Sigma_{GF}$
generated by
\begin{equation}
 \Sigma_{GF}:\quad \bm{\epsilon}_m=
  \epsilon\bm{g}_m\cos\phi+\epsilon\bm{f}_m\sin\phi,
\label{eq:ellipseEpsGF}
\end{equation}
where $0 \le \phi < 2\pi$.
The ellipse $\Sigma_{GF}$ is described by
\begin{equation}
 \left(\Delta g_m\right)^2+\left(\Delta f_m\right)^2
  -2C_{GF}\Delta g_m\Delta f_m
   =1-\left(C_{GF}\right)^2,
\label{eq:SigmaGF}
\end{equation}
with an angle of $-45^\circ$.
The shape of $\Sigma_{GF}$ is characterized by $C_{GF}$,
circular when $C_{GF}=0$,
linear (with slope $-1$) when $C_{GF}=-1$ (see Fig.~\ref{fig5}(b)),
and elliptical (as described above) otherwise.

The points $G^\pm$ and $F^\pm$ correspond to
$\bm{\epsilon}_m=\epsilon\bm{g}^{(\pm)}_m$ and
$\bm{\epsilon}_m=\epsilon\bm{f}^{(\pm)}_m$,
respectively.
Their coordinates are given by
\begin{align}
  G^+: &\quad  \left(1-\delta_{n_0,(d-1)}, 
                              C^{(++)}_{GF}\right), \notag \\
  F^+: &\quad  \left(C^{(++)}_{GF},
                              1-\delta_{n_1,d}\right), \notag \\
  G^-: &\quad  \left(-\cos\theta_g,
                              C^{(-+)}_{GF}\right), \notag \\
  F^-: &\quad  \left(C^{(+-)}_{GF},
                              -\cos\theta_f\right),
\end{align}
where $C^{(-+)}_{GF}=\bm{g}^{(-)}_m\cdot \bm{f}^{(+)}_m\ge0$ and
$C^{(+-)}_{GF}=\bm{g}^{(+)}_m\cdot \bm{f}^{(-)}_m\ge0$ are
given by Eqs.~(\ref{eq:CGF2}) and (\ref{eq:CGF3})
in Appendix~\ref{sec:cosine}.
Note that by Eq.~(\ref{eq:zeroVec1}),
$\bm{g}^{(+)}_m\cdot \bm{g}_m=0$ when $n_0=d-1$
and by Eq.~(\ref{eq:zeroVec2}),
$\bm{f}^{(+)}_m\cdot \bm{f}_m =0$ when $n_1=d$.
The point $G^+$ is the case discussed in the preceding section.

The tilted $\Sigma_{GF}$ indicates that
$\Delta g_m$ and $\Delta f_m$ are negatively correlated.
When $n_1=1$ and $n_0=0$,
$C_{GF}$ can be related to the Pearson correlation coefficient
using isotropic modifications.
That is, let $\bm{\epsilon}^{(n)}_m$ be
a modification normalized to $\epsilon$ for $n=1,2,\ldots,N_p$.
They are assumed to be isotropic as
\begin{align}
  \frac{1}{N_p} \sum_n \epsilon^{(n)}_{mi} &= 0, \\
  \frac{1}{N_p} \sum_n \epsilon^{(n)}_{mi} \epsilon^{(n)}_{mj}
  &= \frac{\epsilon^2}{d}\,\delta_{i,j},
\end{align}
where $\epsilon^{(n)}_{mi}$ is
the $i$th component of $\bm{\epsilon}^{(n)}_m$.
When the points are generated using $\{\bm{\epsilon}^{(n)}_m\}$,
their correlation coefficient is equal to $C_{GF}$.
The perfect negative correlation $C_{GF}=-1$ is achieved by
the optimal measurements $\bm{\lambda}_m=\bm{m}^{(d)}_{1,d-1}(\lambda)$,
as shown in Fig.~\ref{fig5}(b).
Conversely, the non-correlated case $C_{GF}=0$ cannot be achieved
when $n_1=1$ and $n_0=0$.

In contrast, if $n_1>1$ or $n_0>0$,
the measurement $\bm{\lambda}_m$ is on
the relevant boundaries $\lambda_{m1}=\lambda_{mi}$ or $\lambda_{md}=0$,
like the measurement $B$ in Fig.~\ref{fig1}.
Some $\bm{\epsilon}_m$'s are prohibited by
Eqs.~(\ref{eq:condEps1}) and (\ref{eq:condEps2}).
Therefore, the plotted points distribute only in a subregion
of the region enclosed by $\Sigma_{GF}$.
This case is shown in
Fig.~\ref{fig5}(c) for $n_1=2$,
Fig.~\ref{fig5}(d) for $n_0=1$,
and Fig.~\ref{fig5}(e) for $n_1=2$ and $n_0=1$.
Although $G^+$ and $F^+$ are always on $\Sigma_{GF}$,
$G^-$ is not on $\Sigma_{GF}$ if $n_1>1$,
and $F^-$ is not on $\Sigma_{GF}$ if $n_0>0$.

The boundary $\Gamma_{GF}$ of the subregion
consists of four curves
connecting the four points
$G^+$, $F^+$, $G^-$, and $F^-$.
These curves are generated by
\begin{align}
  G^+\to F^+: & \quad
     \bm{\epsilon}_m=
     \epsilon\bm{g}^{(+)}_m\cos\varphi+\epsilon\bm{f}^{(+)}_m\sin\varphi,
       \notag \\
  F^+\to G^-: & \quad
     \bm{\epsilon}_m=
     \epsilon\bm{g}^{(-)}_m\sin\varphi+\epsilon\bm{f}^{(+)}_m\cos\varphi,
       \notag \\
  G^-\to F^-: & \quad
     \bm{\epsilon}_m=
     \epsilon\bm{g}^{(-)}_m\cos\varphi+\epsilon\bm{f}^{(-)}_m\sin\varphi,
       \notag \\
  F^-\to G^+: & \quad
     \bm{\epsilon}_m=
     \epsilon\bm{g}^{(+)}_m\sin\varphi+\epsilon\bm{f}^{(-)}_m\cos\varphi,
\label{eq:arcEpsGF}
\end{align}
where $0 \le \varphi < \pi/2$.
The equations of these curves are provided in Appendix~\ref{sec:arc}.
They are elliptical arcs characterized by
$C^{(++)}_{GF}$, $-C^{(-+)}_{GF}$, $C^{(--)}_{GF}$,
and $-C^{(+-)}_{GF}$,
where $C^{(--)}_{GF}=\bm{g}^{(-)}_m\cdot \bm{f}^{(-)}_m\le0$ is given
by Eq.~(\ref{eq:CGF4}) in Appendix~\ref{sec:cosine}.

Therefore, 
the correlation between $\Delta g_m$ and $\Delta f_m$
can be represented by the four coefficients
$\{C^{(++)}_{GF},-C^{(-+)}_{GF},C^{(--)}_{GF},-C^{(+-)}_{GF}\}$.
If $n_1=1$ and $n_0=0$, the four coefficients are equal.
For example,
$\{-0.60,-0.60,-0.60,-0.60\}$ in Fig.~\ref{fig5}(a)
and $\{-1,-1,-1,-1\}$ in Fig.~\ref{fig5}(b).
Otherwise, the four coefficients are not equal.
In Fig.~\ref{fig5}(c),
they are $\{-0.26,-0.79,-0.79,-0.26\}$,
which denotes that $\Gamma_{GF}$ is flatter
in the left region of the line between $F^+$ and $F^-$
than in the right region.
In Fig.~\ref{fig5}(d),
they are $\{-0.37,-0.37,-0.87,-0.87\}$,
which denotes that $\Gamma_{GF}$ is flatter
in the lower region of the line between $G^+$ and $G^-$
than in the upper region.
In Fig.~\ref{fig5}(e),
$\Gamma_{GF}$ is linear between $G^-$ and $F^-$
as denoted by $\{-0.13,-0.43,-1,-0.32\}$.

Unfortunately,
the case of $\bm{\lambda}_m=\bm{p}^{(d)}_{r}$ is anomalous
in the sense that some of $C^{(\pm\pm)}_{GF}$
fail to characterize $\Gamma_{GF}$.
If $r\neq 1$ and $r\neq d$,
$\Sigma_{GF}$ is a circle from $C_{GF}=0$.
However, $\Gamma_{GF}$ is its first-quadrant quarter,
as shown in Fig.~\ref{fig5}(f),
although the coefficients are $\{0,0,0,0\}$.
If $r=1$ or $r=d$,
those $\Sigma_{GF}$ and $\Gamma_{GF}$ collapse
to lines although $C_{GF}=0$.
The lines are vertical if $r=1$ and horizontal if $r=d$.
The anomalous case occurs because some of the steepest directions
are zero vectors,
as explained in Appendix~\ref{sec:arc}.

%\begin{figure*}
\begin{figure}
\begin{center}
\includegraphics[scale=0.49]{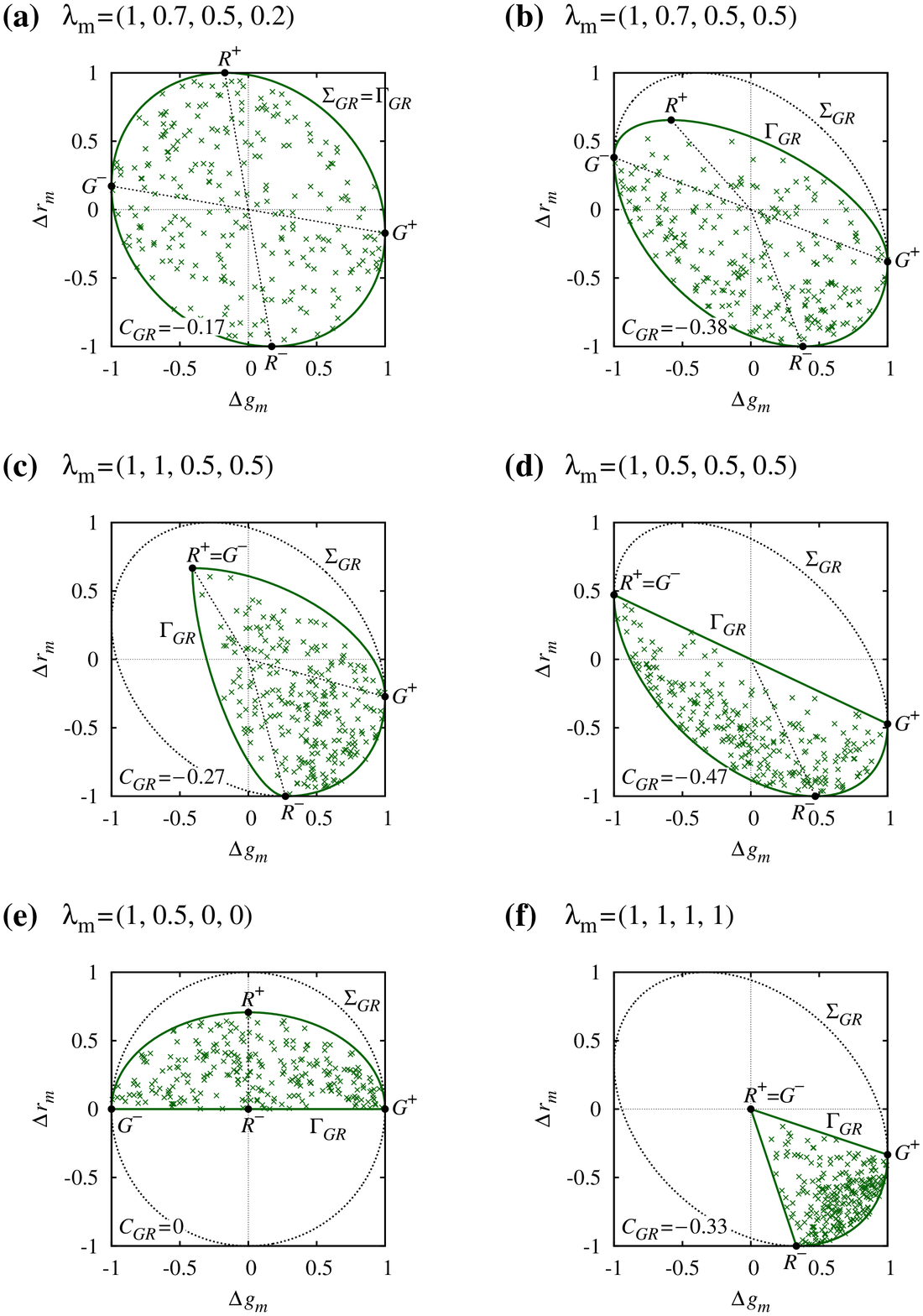}
\end{center}
\caption{\label{fig6}
Correlation between changes in $G(m)$ and $R(m)$.
For six different measurements of $d=4$,
the normalized changes $\Delta g_m$ and $\Delta r_m$
are plotted using $250$ random $\bm{\epsilon}_m$.
The ellipse $\Sigma_{GR}$, characterized by $C_{GR}$, encloses
the region if $\bm{\epsilon}_m$'s were unconditional,
whereas the boundary $\Gamma_{GR}$ encloses
the region when $\bm{\epsilon}_m$'s are
conditioned by Eqs.~(\ref{eq:condEps1})--(\ref{eq:condEps3}).
The points $G^\pm$ and $R^\pm$ are given by
$\bm{\epsilon}_m=\epsilon\bm{g}^{(\pm)}_m$ and
$\bm{\epsilon}_m=\epsilon\bm{r}^{(\pm)}_m$, respectively.
}
\end{figure}%
%\end{figure*}%
Similarly, for $G(m)$ versus $R(m)$,
the change of $R(m)$ in Eq.~(\ref{eq:changeR})
is normalized as
\begin{equation}
  \Delta r_m =\frac{\Delta R(m)}{\left\| \bm{\epsilon}_m \right\|
                                 \left\|\bm{\nabla} R(m) \right\|}
             =\frac{\bm{\epsilon}_m}{\left\| \bm{\epsilon}_m \right\|}
              \cdot \bm{r}_m.
\label{eq:normalChangeR}
\end{equation}
For various $\bm{\epsilon}_m$,
$\Delta g_m$ and $\Delta r_m$ are plotted on a plane.
The modification $\bm{\epsilon}_m$ should satisfy
the conditions of Eqs.~(\ref{eq:condEps1})--(\ref{eq:condEps3}).
Figure~\ref{fig6} shows the plotted graphs
for six measurements of $d=4$.
The points were generated
similarly to the case of $G(m)$ versus $F(m)$.

If $n_1=n_d=1$ and $n_0=0$,
the measurement $\bm{\lambda}_m$ is away from
the relevant boundaries $\lambda_{m1}=\lambda_{mi}$,
$\lambda_{mi}=\lambda_{md}$, and $\lambda_{md}=0$.
It can accept any $\bm{\epsilon}_m$
without constraint from Eqs.~(\ref{eq:condEps1})--(\ref{eq:condEps3}).
Figure~\ref{fig6}(a) shows this case.
The plotted points lie inside the ellipse $\Sigma_{GR}$
described by
\begin{equation}
 \left(\Delta g_m\right)^2+\left(\Delta r_m\right)^2
  -2C_{GR}\Delta g_m\Delta r_m
   =1-\left(C_{GR}\right)^2,
\label{eq:SigmaGR}
\end{equation}
with an angle of $-45^\circ$.
The shape of $\Sigma_{GR}$ is characterized by $C_{GR}$,
although $\Sigma_{GR}$ cannot be linear because $C_{GR}>-1$.
The points $G^\pm$ and $R^\pm$, corresponding to
$\bm{\epsilon}_m=\epsilon\bm{g}^{(\pm)}_m$ and
$\bm{\epsilon}_m=\epsilon\bm{r}^{(\pm)}_m$,
are given by
\begin{align}
  G^+: &\quad  \left(1-\delta_{n_0,(d-1)},
                              -C^{(+-)}_{GR}\right), \notag \\
  R^+: &\quad  \left(C^{(++)}_{GR},
                              \cos\theta_r\right), \notag \\
  G^-: &\quad  \left(-\cos\theta_g,
                              -C^{(--)}_{GR}\right), \notag \\
  R^-: &\quad  \left(C^{(+-)}_{GR},
                              -\delta_{n_0,0}\right),
\end{align}
where $C^{(+-)}_{GR}=\bm{g}^{(+)}_m\cdot \bm{r}^{(-)}_m\ge 0$ and
$C^{(--)}_{GR}=\bm{g}^{(-)}_m\cdot \bm{r}^{(-)}_m\le 0$ are
given by Eqs.~(\ref{eq:CGR3}) and (\ref{eq:CGR4})
in Appendix~\ref{sec:cosine}.

In contrast, unless $n_1=n_d=1$ and $n_0=0$,
the measurement $\bm{\lambda}_m$ is on the relevant boundaries.
Because $\bm{\epsilon}_m$'s are constrained by
Eqs.~(\ref{eq:condEps1})--(\ref{eq:condEps3}),
the plotted points distribute only in a subregion
of the region enclosed by $\Sigma_{GR}$.
This case is shown in Fig.~\ref{fig6}(b) for $n_d=2$,
Fig.~\ref{fig6}(c) for $n_1=n_d=2$,
and Fig.~\ref{fig6}(d) for an optimal measurement $n_d=3$.
Although $G^+$ is always on $\Sigma_{GR}$,
$R^+$ is not on $\Sigma_{GR}$ if $n_d>1$,
$G^-$ is not on $\Sigma_{GR}$ if $n_1>1$,
and $R^-$ is not on $\Sigma_{GR}$ if $n_0>0$.
The boundary $\Gamma_{GR}$ of the subregion consists of
four elliptical arcs connecting the four points
$G^+$, $R^+$, $G^-$, and $R^-$,
as described in Appendix~\ref{sec:arc}.
The arcs are
characterized by $C^{(++)}_{GR}$, $-C^{(-+)}_{GR}$,
$C^{(--)}_{GR}$, and $-C^{(+-)}_{GR}$,
where $C^{(-+)}_{GR}=\bm{g}^{(-)}_m\cdot \bm{r}^{(+)}_m\ge 0$ is given
by Eq.~(\ref{eq:CGR2}) in Appendix~\ref{sec:cosine}.

The correlation between $\Delta g_m$ and $\Delta r_m$
can be represented by the four coefficients
$\{C^{(++)}_{GR},-C^{(-+)}_{GR},C^{(--)}_{GR},-C^{(+-)}_{GR}\}$.
If $n_1=n_d=1$ and $n_0=0$, the coefficients are equal.
For example, they are
$\{-0.17,-0.17,-0.17,-0.17\}$ in Fig.~\ref{fig6}(a).
Otherwise, the coefficients are not equal.
In Fig.~\ref{fig6}(b),
they are $\{-0.58,-0.58,-0.38,-0.38\}$,
which denotes that $\Gamma_{GR}$ is flatter
in the upper region of the line between $G^+$ and $G^-$
than in the lower region.
In Fig.~\ref{fig6}(c),
$R^+$ coincides with $G^-$
as denoted by $\{-0.41,-1,-0.67,-0.27\}$.
In Fig.~\ref{fig6}(d),
$\Gamma_{GR}$ is linear between $G^+$ and $R^+$
for an optimal measurement, as denoted by $\{-1,-1,-0.47,-0.47\}$.

The cases of $\lambda_{md}=0$ and $\bm{\lambda}_m=\bm{p}^{(d)}_{d}$
are anomalous,
as explained in Appendix~\ref{sec:arc}.
In the case of $\lambda_{md}=0$,
$\Sigma_{GR}$ is a circle from $C_{GR}=0$
if $\bm{\lambda}_m\neq\bm{p}^{(d)}_{1}$,
but it collapses to a vertical line
if $\bm{\lambda}_m=\bm{p}^{(d)}_{1}$.
If $\bm{\lambda}_m\neq\bm{p}^{(d)}_{r}$,
$\Gamma_{GR}$ are untilted elliptical arcs
given by Eq.~(\ref{eq:GammaGR2}) in the first quadrant and
Eq.~(\ref{eq:GammaGR3}) in the second quadrant,
but are horizontal lines in the other quadrants (see Fig.~\ref{fig6}(e)),
although the coefficients are $\{0,0,0,0\}$.
Moreover, it collapses to a vertical line in the second quadrant
if $\bm{\lambda}_m=\bm{p}^{(d)}_{r}$, and
likewise in the first quadrant if $\bm{\lambda}_m=\bm{p}^{(d)}_{1}$.
In contrast,
in the case of $\bm{\lambda}_m=\bm{p}^{(d)}_{d}$,
$\Gamma_{GR}$ is an elliptical sector of $\Sigma_{GR}$, as
shown in Fig.~\ref{fig6}(f),
although the coefficients are $\{0,0,0,-1/(d-1)\}$.

\begin{figure}
\begin{center}
\includegraphics[scale=0.52]{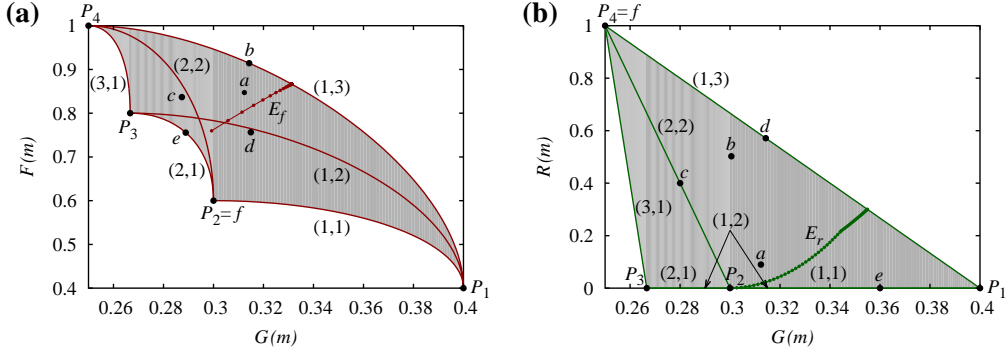}
\end{center}
\caption{\label{fig7}
Allowed regions of
(a) $G(m)$ versus $F(m)$ and (b) $G(m)$ versus $R(m)$.
The point $P_r$ denotes $\bm{p}^{(d)}_{r}$,
and the line $(k,l)$ denotes $\bm{m}^{(d)}_{k,l}(\lambda)$.
The six points from $a$ to $f$ in panel (a)
correspond to the measurements in Fig.~\ref{fig5},
and those in panel (b) correspond to
the measurements in Fig.~\ref{fig6}.
The lines $E_f$ and $E_r$ show
the transitions of measurement as displayed in Fig.~\ref{fig8}
of Sect.~\ref{sec:improvement}.}
\end{figure}%
The local relations shown in
Figs.~\ref{fig5} and \ref{fig6} are consistent with the global ones
shown by the allowed regions of information versus disturbance.
Figure~\ref{fig7}(a) shows
that of $G(m)$ versus $F(m)$~\cite{Terash16}.
Figure~\ref{fig5}(f) reproduces the neighborhood of point $P_2$.
Moreover, Fig.~\ref{fig5}(b) reproduces
the upper boundary $(1,3)$ around the point $b$
except for its lower neighborhood
derived from the higher-order terms in $\bm{\epsilon}_m$,
and Fig.~\ref{fig5}(e) reproduces
the lower boundary $(2,1)$ around the point $e$.
The slopes $\Delta f_m/\Delta g_m=-1$ in Fig.~\ref{fig5}(b) and
$\Delta f_m/\Delta g_m=-\cos\theta_f/\cos\theta_g$ in Fig.~\ref{fig5}(e)
accord with the boundary slope $d F(m)/d G(m)$ in Ref.~\cite{Terash18}
via Eqs.~(\ref{eq:magG}) and (\ref{eq:magF}) in Appendix~\ref{sec:vector}.
Similarly,
as seen from Fig.~\ref{fig7}(b)
showing the allowed region of $G(m)$ versus $R(m)$~\cite{Terash16},
panels (d), (e), and (f) of
Fig.~\ref{fig6} reproduce
the upper boundary $(1,3)$ around the point $d$,
the lower boundary $R(m)=0$ around the point $e$, and
the neighborhood of point $P_4$, respectively.
However, when $\lambda_{md}=0$ like Fig.~\ref{fig6}(e),
$\Delta r_m \neq 0$ does not mean $\Delta R(m)\neq0$,
because $\left\| \bm{\nabla} R(m) \right\|=0$
by Eq.~(\ref{eq:magR}) in Appendix~\ref{sec:vector}.

\section{\label{sec:improvement}Improvability}
Finally, we attempt to improve the measurement
according to the imperfectness of the correlation.
By a general scheme, the measurement is modified
to increase the information extraction
while decreasing the disturbance.
The improvability of the measurement is quantified by
the angle between the steepest-ascent directions
of the information and disturbance.

To improve a measurement $\bm{\lambda}_m$,
the modification $\bm{\epsilon}_m$ should be chosen
such that $\Delta G(m)>0$
with $\Delta F(m)$ or $\Delta R(m)$ also being positive.
The condition for $\bm{\epsilon}_m$ is illustrated
in Fig.~\ref{fig3}(c).
The line $S_I$ orthogonal to $\bm{g}^{(+)}_m$
denotes the surface on which $G(m)$ is constant,
whereas the line $S_D$
orthogonal to $\bm{f}^{(+)}_m$ or $\bm{r}^{(+)}_m$
denotes the surface on which $F(m)$ or $R(m)$ is constant.
If $\bm{\epsilon}_m$ points into the region colored in gray,
the measurement is improved.
Such $\bm{\epsilon}_m$ always exists
unless $\bm{g}^{(+)}_m$ is antiparallel
to $\bm{f}^{(+)}_m$ or $\bm{r}^{(+)}_m$.

For $G(m)$ versus $F(m)$,
the best choice of $\bm{\epsilon}_m$ is
\begin{equation}
 \bm{\epsilon}_m =
   \epsilon\Bigl( \bm{g}^{(+)}_m+\bm{f}^{(+)}_m \Bigr),
\label{eq:improveGF}
\end{equation}
as shown in Fig.~\ref{fig3}(c).
This increases both $G(m)$ and $F(m)$ symmetrically,
\begin{align}
 \Delta G(m) &= \left(1+C^{(++)}_{GF}\right)
                \left[\Delta G(m)\right]_\mathrm{max} \ge 0, \\
 \Delta F(m) &= \left(1+C^{(++)}_{GF}\right)
                \left[\Delta F(m)\right]_\mathrm{max} \ge 0,
\end{align}
from Eqs.~(\ref{eq:maxDelG}) and (\ref{eq:maxDelF}).
Thus, the improvability of the measurement
can be defined by $1+C^{(++)}_{GF}$.
The optimal measurements $\bm{\lambda}_m=\bm{m}^{(d)}_{1,d-1}(\lambda)$
are not improvable because $C^{(++)}_{GF}=-1$, i.e.,
$\bm{g}^{(+)}_m$ is antiparallel to $\bm{f}^{(+)}_m$.
For example, when $C^{(++)}_{GF}=-1$,
no modification $\bm{\epsilon}_m$ obtains
$\Delta G(m)>0$ and $\Delta F(m)>0$, as shown in Fig.~\ref{fig5}(b).

\begin{figure}
\begin{center}
\includegraphics[scale=0.52]{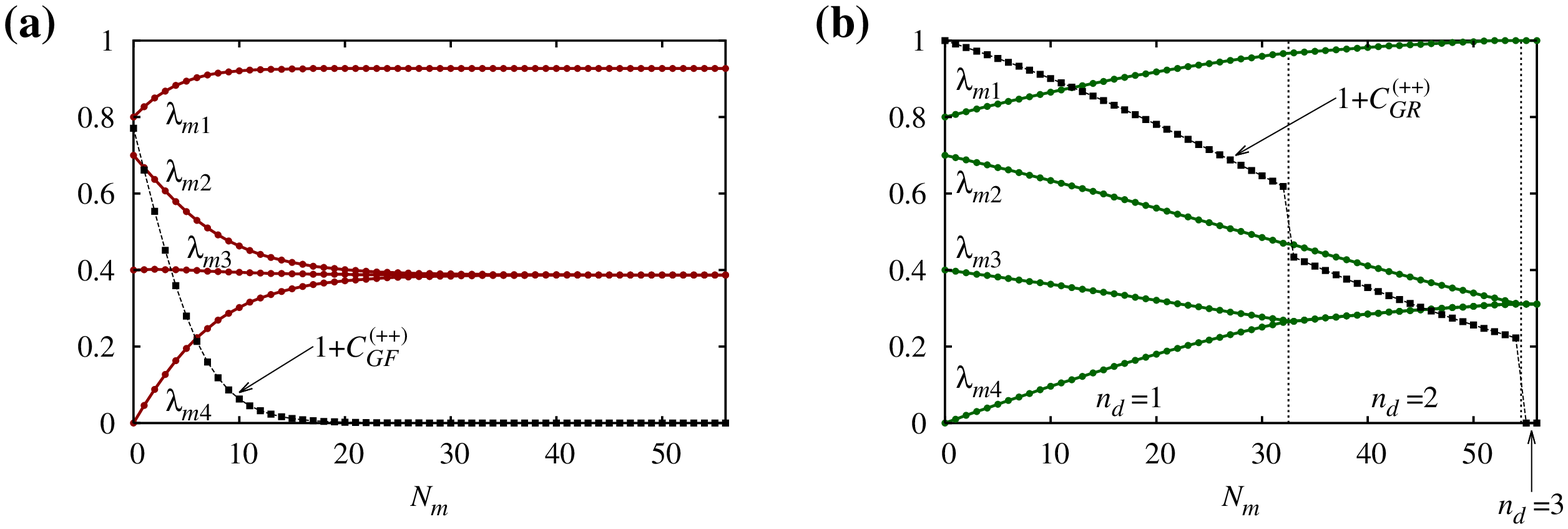}
\end{center}
\caption{\label{fig8}
Transitions of measurement by iterated modifications for
(a) $G(m)$ versus $F(m)$ and (b) $G(m)$ versus $R(m)$.
For the initial measurement
$\bm{\lambda}_m=\left(0.8,0.7,0.4,0\right)$ in $d=4$,
the transitions of singular values $\{\lambda_{mi}\}$ and
improvability $1+C^{(++)}_{GF}$ or $1+C^{(++)}_{GR}$
are shown as functions of the number of modifications $N_m$.}
\end{figure}%
The improvement can be repeated
until the measurement is optimized as $C^{(++)}_{GF}=-1$.
For example, consider improving
a measurement $\bm{\lambda}_m=\left(0.8,0.7,0.4,0\right)$
by iterating Eq.~(\ref{eq:improveGF}) with $\epsilon=0.05$ for $d=4$.
Figure~\ref{fig8}(a) shows the transitions of $\{\lambda_{mi}\}$
and $1+C^{(++)}_{GF}$
as functions of $N_m$, where $N_m$ is the number of modifications.
As $N_m$ increases,
$C^{(++)}_{GF}$ monotonously decreases to $-1$
and exhibits no further change thereafter.
The resultant measurement
$\bm{\lambda}_m=\left(0.93,0.39,0.39,0.39\right)$ is optimal.
As a result,
$G(m)$ increases from $0.30$ to $0.33$,
and $F(m)$ from $0.76$ to $0.87$.
The transitions of $G(m)$ and $F(m)$ are shown
by the line $E_f$ in Fig.~\ref{fig7}(a).

Similarly,
for $G(m)$ versus $R(m)$,
the best choice of $\bm{\epsilon}_m$ is
\begin{equation}
 \bm{\epsilon}_m =
   \epsilon\Bigl( \bm{g}^{(+)}_m+\bm{r}^{(+)}_m \Bigr).
\label{eq:improveGR}
\end{equation}
This increases both $G(m)$ and $R(m)$ symmetrically,
\begin{align}
 \Delta G(m) &= \left(1+C^{(++)}_{GR}\right)
                \left[\Delta G(m)\right]_\mathrm{max} \ge 0, \\
 \Delta R(m) &= \left(1+C^{(++)}_{GR}\right)
                \left[\Delta R(m)\right]_\mathrm{max} \ge 0,
\end{align}
from Eqs.~(\ref{eq:maxDelG}) and (\ref{eq:maxDelR}).
In the case of $G(m)$ versus $R(m)$,
the improvability can be defined by $1+C^{(++)}_{GR}$.
As expected,
the optimal measurements $\bm{\lambda}_m=\bm{m}^{(d)}_{1,d-1}(\lambda)$
are not improvable because $C^{(++)}_{GR}=-1$.
For example, when $C^{(++)}_{GR}=-1$,
no modification $\bm{\epsilon}_m$ obtains
$\Delta G(m)>0$ and $\Delta R(m)>0$, as shown in Fig.~\ref{fig6}(d).

The improvement can be repeated
until the measurement is optimized as $C^{(++)}_{GR}=-1$.
For example, consider improving
the same measurement as the previous example
by iterating Eq.~(\ref{eq:improveGR})
with $\epsilon=0.01$.
Figure~\ref{fig8}(b) shows the transitions of $\{\lambda_{mi}\}$
and $1+C^{(++)}_{GR}$ as functions of $N_m$.
As $N_m$ increases,
$C^{(++)}_{GR}$ monotonously decreases to $-1$,
although it discontinuously decreases at each change in $n_d$.
It exhibits no further change
after it becomes $-1$ with $n_d=d-1$.
During the simulation,
$\bm{\lambda}_m$ was checked if it satisfied Eq.~(\ref{eq:condLamb})
after each modification.
If $\lambda_{m1}>1$,
$\bm{\lambda}_m$ was renormalized to $\lambda_{m1}=1$
by Eq.~(\ref{eq:rescale}) (when $N_m\ge52$).
If $\lambda_{m(4-n_d)}<\lambda_{m4}$,
the last modification was redone using a temporarily reduced $\epsilon$
such that $\lambda_{m(4-n_d)}=\lambda_{m4}$ to update $n_d$
(when $N_m=33,55$).
The resultant measurement
$\bm{\lambda}_m=\left(1,0.31,0.31,0.31\right)$ is optimal.
As a result,
$G(m)$ increases from $0.30$ to $0.35$,
and $R(m)$ from $0$ to $0.30$.
The transitions of $G(m)$ and $R(m)$ are shown
by the line $E_r$ in Fig.~\ref{fig7}(b).

The modifications
in Eqs.~(\ref{eq:improveGF}) and (\ref{eq:improveGR})
slightly change
the probability of the outcome $m$, i.e.,
$p(m)=\sigma_m^2/d$
defined in Eq.~(\ref{eq:totalProb}).
This probability does not change
to first-order in $\bm{\epsilon}_m$
by Eqs.~(\ref{eq:ortho1})--(\ref{eq:ortho3})
provided in Appendix~\ref{sec:vector}.
However, $p(m)$ increases by
the second-order term in $\bm{\epsilon}_m$
and decreases
when $\bm{\lambda}_m$ is rescaled by Eq.~(\ref{eq:rescale}).
In practice,
$p(m)$ increases from $0.32$ to $0.33$ in Fig.~\ref{fig8}(a)
and by less than $0.01$ in Fig.~\ref{fig8}(b).

As a caveat,
the projective measurement $\bm{\lambda}_m=\bm{p}^{(d)}_{1}$
and the identity operation $\bm{\lambda}_m=\bm{p}^{(d)}_{d}$
are exceptions.
They are the singular points of $C^{(++)}_{GF}$ and $C^{(++)}_{GR}$,
for which we have assumed $C^{(++)}_{GF}=C^{(++)}_{GR}=0$.
Therefore,
their improvabilities are calculated to be $1$.
However, they cannot be improved because
they have already reached the maximum $G(m)$, $F(m)$, or $R(m)$.

From the above,
the improvability of a measurement can be quantified
by $1+C^{(++)}_{GF}$ or $1+C^{(++)}_{GR}$.
The measurement can be improved 
as long as the improvability is not zero.
The larger the improvability,
the more effectively the measurement can be improved.
Interestingly, the improvability decreases
in any measurement-improvement process.
This law of improvability decrease is shown in Appendix~\ref{sec:law}.

\section{\label{sec:summary}Summary and Discussion}
We discussed the local trade-off between
information and disturbance in quantum measurements.
When a measurement is slightly modified
to enhance the information,
increased disturbance in the system is the local trade-off.
The measurement was described by the singular values $\bm{\lambda}_m$
of a measurement operator $\hat{M}_m$.
As functions of them,
the information was quantified
by the estimation fidelity $G(m)$,
whereas the disturbance was quantified by
the operation fidelity $F(m)$
and by the physical reversibility $R(m)$.
The present study investigated the local trade-offs
between $G(m)$ and $F(m)$ and between $G(m)$ and $R(m)$.

In the local trade-off,
the directions of steepest ascent and descent of
the information and disturbance play an important role.
For $G(m)$, $F(m)$, and $R(m)$,
their unit vectors in the directions of steepest ascent and descent,
$\bm{g}^{(\pm)}_m$, $\bm{f}^{(\pm)}_m$, and $\bm{r}^{(\pm)}_m$,
were derived from their unit gradient vectors,
$\bm{g}_m$, $\bm{f}_m$, and $\bm{r}_m$.
Using these vectors,
the trade-off was shown as the correlation between
the information and disturbance changes.
The correlation was represented by
the angles between the steepest directions,
$C^{(\pm\pm)}_{GF}=\bm{g}^{(\pm)}_m\cdot \bm{f}^{(\pm)}_m$
or $C^{(\pm\pm)}_{GR}=\bm{g}^{(\pm)}_m\cdot \bm{r}^{(\pm)}_m$.
Moreover, according to the imperfectness of the correlation,
the measurement was improved to enhance the information gain
while diminishing the disturbance.
This improvability was quantified
by $1+C^{(++)}_{GF}$ or $1+C^{(++)}_{GR}$.

The main difference of $R(m)$ from $F(m)$ is
that the steepest-ascent direction $\bm{r}^{(+)}_m$ is not
equal to the gradient direction $\bm{r}_m$
when the minimum singular value degenerates.
This leads to some differences in
the angle, correlation, and improvability.
For example,
the range of the angle $C^{(++)}_{GR}$ is divided into subregions
according to the degeneracy.
In the correlation,
the first elliptical sector characterized by $C^{(++)}_{GR}$
is compressed.
The improvability $1+C^{(++)}_{GR}$ discontinuously
decreases at each change in the degeneracy
during iterated modifications.

Compared to $G(m)$, $F(m)$, and $R(m)$,
their averaged values over outcomes,
$G$, $F$, and $R$~\cite{Banasz01,CheLee12},
are difficult to be analyzed similarly.
Since they are given by $G=\sum_m p(m)G(m)$ and so on,
they are functions of $N_o d$ singular values
when the number of outcomes is $N_o$.
Therefore,
their gradient vectors cannot be made when $N_o$ is indefinite.
Moreover, a measurement modification is difficult to be defined,
because singular values of different outcomes
are not independent of each other by
Eq.~(\ref{eq:completeness}).
Our analysis assumes that
the information and disturbance are
characterized by a fixed number of the same independent parameters.
This implies that
other information--disturbance pairs having such properties
could be analyzed similarly.

The above results are entirely general and fundamental
to the quantum theory of measurements.
They are applicable to
any single-outcome process of an arbitrary measurement.
From the correlation,
there is a trade-off relation within the neighborhood
of the measurement.
This provides a framework
for theorists to develop quantum measurement theories.
From the improvability,
one can find how much the measurement can be improved and how to do it.
This provides a hint for experimentalists to improve
their experiments.
The results in this paper
can broaden our perspectives on quantum measurements,
and are potentially useful for quantum information processing
and communication.

\appendix
\section*{Appendix}

\section{\label{sec:vector}Derivations of Vectors}
Herein, we outline the derivations of the unit vectors
in the gradient, steepest-ascent, and steepest-descent directions
of $G(m)$, $F(m)$, and $R(m)$
under the conditions of Eqs.~(\ref{eq:condEps1})--(\ref{eq:condEps3}).

First, we calculate the unit vectors in the gradient directions
$\bm{g}_m$, $\bm{f}_m$, and $\bm{r}_m$.
The gradient vector of a function $f$ is defined by
\begin{equation}
 \bm{\nabla}f=\left(\frac{\partial f}{\partial \lambda_{m1}},
                \frac{\partial f}{\partial \lambda_{m2}},\ldots,
                \frac{\partial f}{\partial \lambda_{md}}\right).
\end{equation}
From Eqs.~(\ref{eq:Gm})--(\ref{eq:Rm}),
the gradient vectors of $G(m)$, $F(m)$, and $R(m)$
are respectively given by
\begin{align}
  \bm{\nabla} G(m)
       &=\frac{2}{d+1}\left[\frac{\lambda_{m1}}{\sigma_m^2}
       \left(\bm{e}_1
      -\frac{\lambda_{m1}}{\sigma_m^2}\bm{\lambda}_m\right)\right], 
         \label{eq:gradG} \\
  \bm{\nabla} F(m)
       &=\frac{2}{d+1}\left[\frac{\tau_m}{\sigma_m^2}
       \left(\bm{l}_d
      -\frac{\tau_m}{\sigma_m^2}\bm{\lambda}_m\right)\right], \\
  \bm{\nabla} R(m)
       &=2d\left[\frac{\lambda_{md}}{\sigma_m^2}
       \left(\bm{e}_d
      -\frac{\lambda_{md}}{\sigma_m^2}\bm{\lambda}_m\right)\right],
         \label{eq:gradR}
\end{align}
where $\bm{l}_{n}$ is defined by Eq.~(\ref{eq:vecLn}).
Their respective magnitudes are given by
\begin{align}
 \left\| \bm{\nabla} G(m) \right\|
   &=\frac{2}{d+1}\left(\frac{\lambda_{m1}}{\sigma_m^2}
           \sqrt{1-\frac{\lambda_{m1}^2}{\sigma_m^2}}\right),
     \label{eq:magG} \\
 \left\| \bm{\nabla} F(m) \right\|
   &=\frac{2}{d+1}\left(\frac{\tau_m}{\sigma_m^2}
           \sqrt{d-\frac{\tau_m^2}{\sigma_m^2}}\right),
     \label{eq:magF} \\
 \left\| \bm{\nabla} R(m) \right\|
   &=2d\left(\frac{\lambda_{md}}{\sigma_m^2}
           \sqrt{1-\frac{\lambda_{md}^2}{\sigma_m^2}}\right).
     \label{eq:magR}
\end{align}
By dividing the gradient vectors by their magnitudes,
the unit vectors in the gradient directions are
given as Eqs.~(\ref{eq:unitG})--(\ref{eq:unitR}).

\begin{figure}
\begin{center}
\includegraphics[scale=0.42]{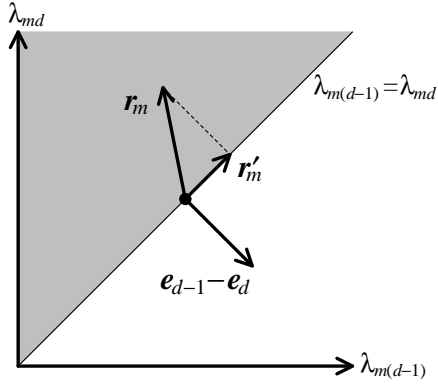}
\end{center}
\caption{\label{fig9}
Gradient vector of $R(m)$ on boundary.
The vector $\bm{r}_m$ is the unit gradient vector of $R(m)$
for a measurement on the boundary $\lambda_{m(d-1)}=\lambda_{md}$.
The gray region shows the forbidden region $\lambda_{m(d-1)}<\lambda_{md}$,
and the vector $\bm{e}_{d-1}-\bm{e}_d$ is normal to the boundary.
By projecting $\bm{r}_m$ onto the boundary,
the vector $\bm{r}'_{m}$ is obtained.}
\end{figure}%
Next, we consider the unit vectors in the steepest-ascent directions
$\bm{g}^{(+)}_m$, $\bm{f}^{(+)}_m$, and $\bm{r}^{(+)}_m$.
They are not necessarily equal to
$\bm{g}_m$, $\bm{f}_m$, and $\bm{r}_m$ under
the conditions of Eqs.~(\ref{eq:condEps1})--(\ref{eq:condEps3}).
$\bm{\epsilon}_m=\epsilon\bm{g}_m$ and
$\bm{\epsilon}_m=\epsilon\bm{f}_m$
with a positive infinitesimal $\epsilon$ satisfy
the conditions.
This means $\bm{g}^{(+)}_m=\bm{g}_m$ and $\bm{f}^{(+)}_m=\bm{f}_m$
in Eqs.~(\ref{eq:ascG}) and (\ref{eq:ascF}).
However, $\bm{\epsilon}_m=\epsilon\bm{r}_m$
violates the condition of Eq.~(\ref{eq:condEps3})
if the minimum singular value degenerates:
\begin{equation}
 \lambda_{m(d-n_d+1)}=\lambda_{m(d-n_d+2)}=\cdots=\lambda_{md},
\end{equation}
where $n_d$ is the degeneracy of the minimum singular value.
For example, if $\lambda_{m(d-1)}=\lambda_{md}$ and $i=d-1$,
the condition of Eq.~(\ref{eq:condEps3}) is violated as
\begin{equation}
 \left(\bm{e}_{d-1}-\bm{e}_d\right) \cdot \bm{r}_m
   =-\frac{\sigma_m}{\sqrt{\sigma_m^2-\lambda_{md}^2}} < 0.
\end{equation}
In this case, $\bm{r}_m$ points
from the boundary $\lambda_{m(d-1)}=\lambda_{md}$
into the forbidden region $\lambda_{m(d-1)}<\lambda_{md}$
(see Fig.~\ref{fig9}).
In the forbidden region,
$\bm{\lambda}'_m= \bm{\lambda}_m +\bm{\epsilon}_m$
should be rearranged such that $\lambda'_{md}$ is
smaller than the other singular values,
meaning that $R(m)$ decreases rather than increases.
For example,
when $d=3$ and $\bm{\lambda}_m=\left(1,1/2,1/2\right)$,
$\bm{\epsilon}_m=\epsilon\bm{r}_m$ generates
\begin{align}
\bm{\lambda}'_m 
    &=\left(1-\frac{2\epsilon}{\sqrt{30}},
            \frac{1}{2}-\frac{\epsilon}{\sqrt{30}},
            \frac{1}{2}+\frac{5\epsilon}{\sqrt{30}}\right) \notag \\
    & \overset{\text{After rearranging}}{\longrightarrow}
      \left(1-\frac{2\epsilon}{\sqrt{30}},
            \frac{1}{2}+\frac{5\epsilon}{\sqrt{30}},
            \frac{1}{2}-\frac{\epsilon}{\sqrt{30}}\right),
\end{align}
$R(m)$ decreases from $1/2$ to
$\left(1/2\right)-\left(2\epsilon/\sqrt{30}\right)$.
Hence, $\bm{r}_m$ does not point
in the steepest-ascent direction of $R(m)$
if $\lambda_{m(d-1)}=\lambda_{md}$.

If $\lambda_{m(d-1)}=\lambda_{md}$,
the steepest-ascent direction of $R(m)$ is given by
the vector $\bm{r}'_{m}$
obtained by projecting $\bm{r}_m$ onto the boundary,
as shown in Fig.~\ref{fig9}.
Using the unit normal vector of the boundary,
\begin{equation}
\bm{n}=\frac{1}{\sqrt{2}}\left(\bm{e}_{d-1}-\bm{e}_{d}\right),
\end{equation}
the projected vector is given by
\begin{equation}
  \bm{r}'_{m}
   =\frac{\sigma_m}{\sqrt{\sigma_m^2-\lambda_{md}^2}}
       \left(\frac{\bm{e}_{d-1}+\bm{e}_d}{2}
      -\frac{\lambda_{md}}{\sigma_m^2}\bm{\lambda}_m\right),
\end{equation}
which satisfies the condition of Eq.~(\ref{eq:condEps3}) when $i=d-1$.
However,
if $\lambda_{m(d-2)}=\lambda_{md}$,
the condition of Eq.~(\ref{eq:condEps3}) is violated when $i=d-2$.
In that case,
$\bm{r}'_{m}$ is again projected onto the boundary
$\lambda_{m(d-2)}=\lambda_{md}$ using the unit normal vector
\begin{equation}
\bm{n}'=\sqrt{\frac{2}{3}}\left(\bm{e}_{d-2}-
       \frac{\bm{e}_{d-1}+\bm{e}_d}{2}\right).
\end{equation}
In general,
if the degeneracy of the minimum singular value is $n_d$,
$\bm{r}_m$ should be projected $(n_d-1)$ times
to satisfy the condition of Eq.~(\ref{eq:condEps3}) for all $i$.
After normalizing the projected vector,
the unit vector in the steepest-ascent direction
of $R(m)$ is determined as Eq.~(\ref{eq:ascR}).

Finally, we consider the unit vectors in the steepest-descent directions
$\bm{g}^{(-)}_m$, $\bm{f}^{(-)}_m$, and $\bm{r}^{(-)}_m$.
They are not necessarily equal to
$-\bm{g}_m$, $-\bm{f}_m$, and $-\bm{r}_m$ under
the conditions of Eqs.~(\ref{eq:condEps1})--(\ref{eq:condEps3}).
For example,
$\bm{\epsilon}_m=-\epsilon\bm{g}_m$
violates the condition of Eq.~(\ref{eq:condEps2})
if the maximum singular value degenerates:
\begin{equation}
 \lambda_{m1}=\lambda_{m2}=\cdots=\lambda_{mn_1},
\end{equation}
where $n_1$ is the degeneracy of the maximum singular value.
By projecting and normalizing $-\bm{g}_m$,
the unit vector in the steepest-descent direction
of $G(m)$ is given as Eq.~(\ref{eq:dscG}).
Similarly,
$\bm{\epsilon}_m=-\epsilon\bm{f}_m$
violates the condition of Eq.~(\ref{eq:condEps1})
if some singular values are $0$:
\begin{equation}
 \lambda_{m(d-n_0+1)}=\lambda_{m(d-n_0+2)}=\cdots=\lambda_{md}=0,
\end{equation}
where $n_0$ is the degeneracy of the singular value $0$.
$n_0=0$ if $\lambda_{md}\neq 0$,
and $n_0=n_d$ if $\lambda_{md}=0$.
By projecting and normalizing $-\bm{f}_m$,
the unit vector in the steepest-descent direction
of $F(m)$ is given as Eq.~(\ref{eq:dscF}).
Moreover,
$\bm{\epsilon}_m=-\epsilon\bm{r}_m$
violates the condition of Eq.~(\ref{eq:condEps1})
if $\lambda_{md}=0$,
because $-\bm{r}_m=-\bm{e}_d$.
As the projected vector of $-\bm{e}_d$
on the boundary $\lambda_{md}=0$ is the zero vector $\bm{0}$,
the unit vector in the steepest-descent direction
of $R(m)$ is $-\bm{r}_m$ if $\lambda_{md}\neq0$ but
$\bm{0}$ if $\lambda_{md}=0$.
This is equivalently written as Eq.~(\ref{eq:dscR}),
because $n_0= 0$ if $\lambda_{md}\neq0$,
and $n_0\neq 0$ if $\lambda_{md}=0$.

All of the derived vectors
are orthogonal to $\bm{\lambda}_m$:
\begin{align}
 \bm{\lambda}_m\cdot \bm{g}_m =
 \bm{\lambda}_m\cdot \bm{g}^{(\pm)}_m &= 0, \label{eq:ortho1} \\
 \bm{\lambda}_m\cdot \bm{f}_m =
 \bm{\lambda}_m\cdot \bm{f}^{(\pm)}_m &= 0, \\
 \bm{\lambda}_m\cdot \bm{r}_m =
 \bm{\lambda}_m\cdot \bm{r}^{(\pm)}_m &= 0. \label{eq:ortho3} 
\end{align}
This originates from the invariance of $G(m)$, $F(m)$, and $R(m)$ 
under the rescaling operation in Eq.~(\ref{eq:rescale}).

\section{\label{sec:cosine}Formulas for Angles}
Herein, we show the angles
between the steepest directions of the information and disturbance.
Two information--disturbance pairs are discussed:
$G(m)$ versus $F(m)$ and $G(m)$ versus $R(m)$.

For $G(m)$ versus $F(m)$,
the cosines of the angles are defined by
$C^{(\pm\pm)}_{GF}=\bm{g}^{(\pm)}_m\cdot \bm{f}^{(\pm)}_m$ and
$C_{GF}=\bm{g}_m\cdot \bm{f}_m$.
$C^{(++)}_{GF}$ and $C_{GF}$ are given by Eqs.~(\ref{eq:CGF})
and (\ref{eq:CGF0}), respectively.
The remaining angles are
\begin{align}
 C^{(-+)}_{GF} &=
 \frac{\sqrt{n_1}\left(\tau_m\lambda_{m1}-\sigma_m^2\right)}
             {\sqrt{\left(\sigma_m^2-n_1\lambda_{m1}^2\right)
                    \left(d\sigma_m^2-\tau_m^2\right)}} 
    \ge 0,  \label{eq:CGF2} \\
 C^{(+-)}_{GF} &=
 \frac{\tau_m\lambda_{m1}-\sigma_m^2}
             {\sqrt{\left(\sigma_m^2-\lambda_{m1}^2\right)
                    \left[\left(d-n_0\right)\sigma_m^2-\tau_m^2\right]}}
   \ge 0,   \label{eq:CGF3} \\
 C^{(--)}_{GF} &= 
 -\frac{\sqrt{n_1}\left(\tau_m\lambda_{m1}-\sigma_m^2\right)}
             {\sqrt{\left(\sigma_m^2-n_1\lambda_{m1}^2\right)
                    \left[\left(d-n_0\right)\sigma_m^2-\tau_m^2\right]}}
    \le 0. \label{eq:CGF4}
\end{align}
Therefore, the angles between $\bm{g}^{(-)}_m$ and $\bm{f}^{(+)}_m$,
and $\bm{g}^{(+)}_m$ and $\bm{f}^{(-)}_m$
are either acute or right,
whereas the angle between $\bm{g}^{(-)}_m$ and $\bm{f}^{(-)}_m$
is either right or obtuse.
Using Eqs.~(\ref{eq:cosG}) and (\ref{eq:cosF}),
all the cosines are related as follows:
\begin{align}
 & C^{(++)}_{GF}
   =-C^{(-+)}_{GF}\cos\theta_g=C^{(--)}_{GF}\cos\theta_g\cos\theta_f \notag \\
 & \qquad\qquad
   =-C^{(+-)}_{GF}\cos\theta_f=C_{GF}.
\label{eq:relCGF}
\end{align}

Similarly,
for $G(m)$ versus $R(m)$,
the cosines of the angles are defined by
$C^{(\pm\pm)}_{GR}=\bm{g}^{(\pm)}_m\cdot \bm{r}^{(\pm)}_m$ and
$C_{GR}=\bm{g}_m\cdot \bm{r}_m$.
$C^{(++)}_{GR}$ and $C_{GR}$ are
given by Eqs.~(\ref{eq:CGR}) and (\ref{eq:CGR0}), respectively.
The remaining angles are
\begin{align}
 C^{(-+)}_{GR} &= 
  \frac{\sqrt{n_1n_d}\,\lambda_{m1}\lambda_{md}
                \left(1-\delta_{n_d,d}\right)}
             {\sqrt{\left(\sigma_m^2-n_1\lambda_{m1}^2\right)
                    \left(\sigma_m^2-n_d\lambda_{md}^2\right)}}
    \ge 0, \label{eq:CGR2} \\
 C^{(+-)}_{GR} &=-C_{GR} \ge 0, \label{eq:CGR3} \\
 C^{(--)}_{GR} &=
    -\frac{\sqrt{n_1}\,\lambda_{m1}\lambda_{md}
                \left(1-\delta_{n_1,d}\right)}
             {\sqrt{\left(\sigma_m^2-n_1\lambda_{m1}^2\right)
                    \left(\sigma_m^2-\lambda_{md}^2\right)}}
    \le 0. \label{eq:CGR4}
\end{align}
The angles between $\bm{g}^{(-)}_m$ and $\bm{r}^{(+)}_m$,
and $\bm{g}^{(+)}_m$ and $\bm{r}^{(-)}_m$
are either acute or right,
whereas the angle between $\bm{g}^{(-)}_m$ and $\bm{r}^{(-)}_m$
is either right or obtuse.
Using Eqs.~(\ref{eq:cosR}) and (\ref{eq:cosG}),
all the cosines are related as follows:
\begin{align}
 & C^{(++)}_{GR}\cos\theta_r
      =-C^{(-+)}_{GR}\cos\theta_g\cos\theta_r \notag \\
 & \qquad\qquad
   =C^{(--)}_{GR}\cos\theta_g=-C^{(+-)}_{GR}=C_{GR}
\label{eq:relCGR}
\end{align}
if $\bm{\lambda}_m\neq\bm{p}^{(d)}_{d}$.
In contrast, if $\bm{\lambda}_m=\bm{p}^{(d)}_{d}$,
the third equality in Eq.~(\ref{eq:relCGR}) does not hold,
because $C^{(++)}_{GR}=C^{(-+)}_{GR}=C^{(--)}_{GR}=0$
but $-C^{(+-)}_{GR}=C_{GR}=-1/(d-1)$.

\section{\label{sec:arc}Equations of Arcs}
Herein, we describe the boundary equation in which
the normalized changes of the information and disturbance
are contained.
The boundary consists of four elliptical arcs
characterized by the angles between the steepest directions.

For $G(m)$ versus $F(m)$,
the normal case $\bm{\lambda}_m\neq\bm{p}^{(d)}_{r}$
is first considered.
The boundary $\Gamma_{GF}$ is generated by Eq.~(\ref{eq:arcEpsGF}).
Between $G^+$ and $F^+$, it coincides with the ellipse $\Sigma_{GF}$
from Eqs.~(\ref{eq:ascG}) and (\ref{eq:ascF}).
Therefore, the arc in this interval is
described by Eq.~(\ref{eq:SigmaGF}),
characterized by $C^{(++)}_{GF}$ from Eq.~(\ref{eq:CGF0}).
However, between $F^+$ and $G^-$,
$\Gamma_{GF}$ is an elliptical arc described by
\begin{equation}
 \left(\Delta g'_m\right)^2+\left(\Delta f_m\right)^2
  +2C^{(-+)}_{GF}\Delta g'_m\Delta f_m
   =1-\left[C^{(-+)}_{GF}\right]^2,
\label{eq:GammaGF2}
\end{equation}
where we have used
$\Delta g'_m=\Delta g_m/\cos\theta_g$ and Eq.~(\ref{eq:relCGF}).
This ellipse is obtained from $\Sigma_{GF}$
by replacing $C_{GF}$ with $-C^{(-+)}_{GF}=C_{GF}/\cos\theta_g$
and horizontally compressing by a factor of $1/\cos\theta_g$.
The compression is just for the arc to be connected
with the adjacent arcs at $F^+$ and $G^-$.
Thus, the arc in this interval is
characterized by $-C^{(-+)}_{GF}$.
When $C^{(-+)}_{GF}=0$,
the ellipse in Eq.~(\ref{eq:GammaGF2}) is untilted
(with axes $\cos\theta_g$ and $1$),
but when $C^{(-+)}_{GF}=1$, 
it collapses to a line (with slope $-1/\cos\theta_g$).
In the latter case, $F^+$ coincides with $G^-$,
which means that
the arc in this interval shrinks to a point.

Moreover,
$\Gamma_{GF}$ is an elliptical arc described by
\begin{equation}
 \left(\Delta g'_m\right)^2+\left(\Delta f'_m\right)^2
  -2C^{(--)}_{GF}\Delta g'_m\Delta f'_m
   =1-\left[C^{(--)}_{GF}\right]^2
\end{equation}
with
$\Delta f'_m=\Delta f_m/\cos\theta_f$ between $G^-$ and $F^-$,
and is an elliptical arc described by
\begin{equation}
 \left(\Delta g_m\right)^2+\left(\Delta f'_m\right)^2
  +2C^{(+-)}_{GF}\Delta g_m\Delta f'_m
   =1-\left[C^{(+-)}_{GF}\right]^2
\end{equation}
between $F^-$ and $G^+$.
The arcs in these intervals are
characterized by $C^{(--)}_{GF}$ and $-C^{(+-)}_{GF}$,
respectively.
When $C^{(--)}_{GF}=-1$, $\Gamma_{GF}$ is linear
between $G^-$ and $F^-$, as shown in Fig.~\ref{fig5}(e),
and when $C^{(+-)}_{GF}=1$, $F^-$ coincides with $G^+$.

In contrast,
the case of $\bm{\lambda}_m=\bm{p}^{(d)}_{r}$ is anomalous
in the sense that some of $C^{(\pm\pm)}_{GF}$
fail to characterize $\Gamma_{GF}$.
This is because $\bm{g}^{(-)}_m=\bm{f}^{(-)}_m = \bm{0}$
from Eq.~(\ref{eq:zeroVec3}).
The elliptical arcs generated by Eq.~(\ref{eq:arcEpsGF})
collapse to lines or points regardless of $C^{(\pm\pm)}_{GF}=0$
except for the first one.
In addition, $\bm{g}_m=\bm{g}^{(+)}_m = \bm{0}$
if $r=1$ from Eq.~(\ref{eq:zeroVec1}),
and $\bm{f}_m=\bm{f}^{(+)}_m= \bm{0}$
if $r=d$ from Eq.~(\ref{eq:zeroVec2}).
These also collapse
the first arc generated by Eq.~(\ref{eq:arcEpsGF})
and $\Sigma_{GF}$ generated by Eq.~(\ref{eq:ellipseEpsGF})
to vertical or horizontal lines.
The explicit shape of $\Gamma_{GF}$ 
in the anomalous case is described
in the main text.

Similarly,
for $G(m)$ versus $R(m)$,
the normal case,
$\lambda_{md}\neq0$ and $\bm{\lambda}_m\neq\bm{p}^{(d)}_{d}$,
is first considered.
The boundary $\Gamma_{GR}$ is generated by
a similar equation to Eq.~(\ref{eq:arcEpsGF}),
but using $\bm{r}^{(\pm)}_m$ instead of $\bm{f}^{(\pm)}_m$.
Between $G^+$ and $R^+$,
$\Gamma_{GR}$ is an elliptical arc
described by
\begin{equation}
 \left(\Delta g_m\right)^2+\left(\Delta r'_m\right)^2
  -2C^{(++)}_{GR}\Delta g_m\Delta r'_m
   =1-\left[C^{(++)}_{GR}\right]^2,
\label{eq:GammaGR2}
\end{equation}
where we have used $\Delta r'_m=\Delta r_m/\cos\theta_r$
and Eq.~(\ref{eq:relCGR}).
This ellipse is obtained from $\Sigma_{GR}$
by replacing $C_{GR}$ with $C^{(++)}_{GR}=C_{GR}/\cos\theta_r$
and vertically compressing by a factor of $1/\cos\theta_r$.
Thus, the arc in this interval is
characterized by $C^{(++)}_{GR}$.
When $C^{(++)}_{GR}=0$,
the ellipse in Eq.~(\ref{eq:GammaGR2}) is untilted
(with axes $1$ and $\cos\theta_r$),
but when $C^{(++)}_{GR}=-1$,
it collapses to a line (with slope $-\cos\theta_r$).
In the latter case,
$\Gamma_{GR}$ is linear between $G^+$ and $R^+$,
as shown in Fig.~\ref{fig6}(d).

Moreover,
$\Gamma_{GR}$ is an elliptical arc described by
\begin{equation}
 \left(\Delta g'_m\right)^2+\left(\Delta r'_m\right)^2
  +2C^{(-+)}_{GR}\Delta g'_m\Delta r'_m
   =1-\left[C^{(-+)}_{GR}\right]^2
   \label{eq:GammaGR3}
\end{equation}
between $R^+$ and $G^-$,
and is an elliptical arc described by
\begin{equation}
 \left(\Delta g'_m\right)^2+\left(\Delta r_m\right)^2
  -2C^{(--)}_{GR}\Delta g'_m\Delta r_m
   =1-\left[C^{(--)}_{GR}\right]^2
\end{equation}
between $G^-$ and $R^-$.
The arcs in these intervals are
characterized by $-C^{(-+)}_{GR}$ and $C^{(--)}_{GR}$,
respectively.
For example,
when $C^{(-+)}_{GR}=1$, $R^+$ coincides with $G^-$,
as shown in Fig.~\ref{fig6}(c),
and when $C^{(--)}_{GR}=-1$,
$\Gamma_{GR}$ is linear between $G^-$ and $R^-$.
Finally, $\Gamma_{GR}$
coincides with $\Sigma_{GR}$ between $R^-$ and $G^+$.
The arc in this interval is
described by Eq.~(\ref{eq:SigmaGR}),
characterized by $-C^{(+-)}_{GR}$ from Eq.~(\ref{eq:CGR3}).

However,
the cases of $\lambda_{md}=0$ and
$\bm{\lambda}_m=\bm{p}^{(d)}_{d}$ are anomalous.
In the case of $\lambda_{md}=0$,
$\bm{r}^{(-)}_m = \bm{0}$ from Eq.~(\ref{eq:dscR}).
This collapses
the third and fourth arcs of $\Gamma_{GR}$ to horizontal lines.
Moreover, $\bm{g}^{(-)}_m = \bm{0}$
if $\bm{\lambda}_m=\bm{p}^{(d)}_{r}$ from Eq.~(\ref{eq:zeroVec3}),
collapsing the second arc to a vertical line.
In addition, $\bm{g}_m=\bm{g}^{(+)}_m = \bm{0}$
if $\bm{\lambda}_m=\bm{p}^{(d)}_{1}$ from Eq.~(\ref{eq:zeroVec1}),
collapsing
the first arc and $\Sigma_{GR}$ also collapse to vertical lines.
In contrast, in the case of $\bm{\lambda}_m=\bm{p}^{(d)}_{d}$,
$\bm{r}^{(+)}_m =\bm{g}^{(-)}_m = \bm{0}$
from Eqs.~(\ref{eq:zeroVec2}) and (\ref{eq:zeroVec3}).
The first and third arcs of $\Gamma_{GR}$ collapse
to lines
tilting by $C^{(+-)}_{GR}\neq 0$
and the second arc to a point.
The explicit shapes of $\Gamma_{GR}$ 
in the anomalous cases are described
in the main text.

\section{\label{sec:law}Law of Improvability Decrease}
Herein, we outline the proof of the law of improvability decrease.
It states that the improvability decreases
in any measurement-improvement process.
That is, an improved measurement is always less improvable than
the original measurement.

For $G(m)$ versus $F(m)$,
suppose that a measurement $\bm{\lambda}_m$ is modified by
an arbitrary $\bm{\epsilon}_m$.
This modification changes $\lambda_{m1}$, $\tau_m$, and $\sigma_m^2$
by $\Delta\lambda_{m1}$, $\Delta \tau_m$, and $\Delta \sigma_m^2$,
respectively.
To first-order in these changes,
$\Delta G(m)$ and $\Delta F(m)$ are expanded as
\begin{align}
 \Delta G(m) &= \frac{2}{d+1}\left[\frac{\lambda_{m1}}{\sigma_m^2}
       \left(\Delta\lambda_{m1}
      -\frac{\lambda_{m1}}{2\sigma_m^2}\Delta \sigma_m^2 \right)\right], 
     \label{eq:expandG} \\
 \Delta F(m) &= \frac{2}{d+1}\left[\frac{\tau_m}{\sigma_m^2}
       \left(\Delta \tau_m
      -\frac{\tau_m}{2\sigma_m^2}\Delta \sigma_m^2\right)\right].
     \label{eq:expandF}
\end{align}
$\Delta C^{(++)}_{GF}$ can be expanded similarly.
By eliminating $\Delta\lambda_{m1}$, $\Delta \tau_m$, and $\Delta \sigma_m^2$
from $\Delta C^{(++)}_{GF}$
using Eqs.~(\ref{eq:expandG}) and (\ref{eq:expandF}),
$\Delta C^{(++)}_{GF}$ is related to
$\Delta G(m)$ and $\Delta F(m)$ as
\begin{align}
\Delta C^{(++)}_{GF}
   &=\frac{(d+1)\sigma_m^4}{2\left(\tau_m\lambda_{m1}-\sigma_m^2\right)}
   \left[ \frac{\tau_m-\lambda_{m1}}
         {\lambda_{m1}\left(\sigma_m^{2}-\lambda_{m1}^2\right)}
          \Delta G(m) \right. \notag \\ 
    & \qquad\qquad\quad{}+ 
    \left.\frac{d\lambda_{m1}-\tau_m}
    {\tau_m\left(d\sigma_m^2-\tau_m^2\right)} \Delta F(m)\right]C^{(++)}_{GF}.
\end{align}
This means that $\Delta C^{(++)}_{GF}<0$
if $\Delta G(m)>0$ and $\Delta F(m)>0$,
proving the law of improvability decrease.

In contrast,
for $G(m)$ versus $R(m)$,
it suffices to consider that $\bm{\epsilon}_m$ does not change $n_d$.
This is because when $n_d$ is increased by reaching the boundary,
$C^{(++)}_{GR}$ decreases,
whereas when $n_d$ is decreased by leaving the boundary,
$G(m)$ and $R(m)$ cannot increase simultaneously
(see Fig.~\ref{fig2}).
In terms of $\Delta\lambda_{m1}$, $\Delta\lambda_{md}$,
and $\Delta \sigma_m^2$,
$\Delta R(m)$ is expanded as
\begin{equation}
 \Delta R(m) = 2d\left[\frac{\lambda_{md}}{\sigma_m^2}
       \left(\Delta\lambda_{md}
      -\frac{\lambda_{md}}{2\sigma_m^2}\Delta \sigma_m^2\right)\right].
\label{eq:expandR}
\end{equation}
$\Delta C^{(++)}_{GR}$ can be expanded similarly.
By eliminating $\Delta\lambda_{m1}$, $\Delta\lambda_{md}$,
and $\Delta \sigma_m^2$
from $\Delta C^{(++)}_{GR}$
using Eqs.~(\ref{eq:expandG}) and (\ref{eq:expandR}),
$\Delta C^{(++)}_{GR}$ is related to $\Delta G(m)$ and $\Delta R(m)$ as
\begin{align}
\Delta C^{(++)}_{GR}
    & =\frac{\sigma_m^4}{2}
   \left[\frac{d+1}{\lambda_{m1}^2\left(\sigma_m^{2}-\lambda_{m1}^2\right)}
          \Delta G(m) \right. \notag \\ 
    & \qquad\quad{}+ \left.
      \frac{1}{d\lambda_{md}^2\left(\sigma_m^{2}-n_d\lambda_{md}^2\right)}
      \Delta R(m)\right]C^{(++)}_{GR}.
\end{align}
This means that $\Delta C^{(++)}_{GR}<0$
if $\Delta G(m)>0$ and $\Delta R(m)>0$,
proving the law of improvability decrease.

%\bibliographystyle{prsty}
%\bibliography{art,bk,paper}
%\end{document}

\end{document}